\documentclass[11pt]{article}

\pdfoutput=1

\usepackage[makeroom]{cancel}
\usepackage{color}
\usepackage{graphicx}
\usepackage{amsmath}
\usepackage{amssymb}
\usepackage{xspace}
\usepackage[small]{subfigure}
\usepackage[numbers,compress]{natbib}
\usepackage[hyperfootnotes=false]{hyperref}

\newlength{\fighskip} \fighskip=2pt
\newlength{\figvskip} \figvskip=3pt

\newcommand*{\figbox}[2]{{
  \def\figscale{#1}
  \def\arraystretch{0.8}
  \arraycolsep=0pt
  \begin{array}{c}
    \vbox{\vskip\figscale\figvskip
      \hbox{\hskip\figscale\fighskip
        \includegraphics[scale=\figscale]{#2}}}
  \end{array}}}

\usepackage{mciteplus} 
\usepackage{dcolumn}
\usepackage{bm}
\usepackage{verbatim}
\usepackage{amscd}
\usepackage{amsfonts}
\usepackage{setspace}
\usepackage{amsthm}
\usepackage{enumerate}
\usepackage{mathtools}

\theoremstyle{plain}

\theoremstyle{plain}

\theoremstyle{plain}
 
\theoremstyle{plain}

\theoremstyle{remark}

\theoremstyle{conjecture}

\theoremstyle{observation}

\theoremstyle{definition}

\theoremstyle{corollary}

\theoremstyle{definition}

\theoremstyle{definition}

\theoremstyle{result}

\theoremstyle{assumption}

\theoremstyle{definition}

\theoremstyle{problem}

\theoremstyle{fact}

\DeclareMathOperator{\Tr}{Tr}

\begin{document}

\title{\bf 
Firewalls vs. Scrambling}
\author{
Beni Yoshida\\ 
{\em \small Perimeter Institute for Theoretical Physics, Waterloo, Ontario N2L 2Y5, Canada} }
\date{}

\maketitle

\begin{abstract}

Recently we pointed out that the black hole interior operators can be reconstructed by using the Hayden-Preskill recovery protocols. Building on this observation, we propose a resolution of the firewall problem by presenting a state-independent reconstruction of interior operators. 
Our construction avoids the non-locality problem which plagued the ``$A=R_{B}$'' or ``$\text{ER}=\text{EPR}$'' proposals. 
We show that the gravitational backreaction by the infalling observer, who simply falls into a black hole, disentangles the outgoing mode from the early radiation.
The infalling observer crosses the horizon smoothly and sees quantum entanglement between the outgoing mode and the interior mode which is distinct from the originally entangled qubit in the early radiation. 
Namely, quantum operation on the early radiation cannot influence the experience of the infalling observer since description of the interior mode does not involve the early radiation at all. 
We also argue that verification of quantum entanglement by the outside observer does not create a firewall. Instead it will perform the Hayden-Preskill recovery which saves an infalling observer from crossing the horizon. 
\end{abstract}

\tableofcontents
\newpage

\section{Introduction}

The question of how a quantum black hole forms and evaporates unitarily remains among the most puzzling mysteries in theoretical physics~\cite{Hawking:1976aa}. Following earlier works, Almheiri, Marolf, Polchinski and Sully (AMPS) argued that by careful quantum processing of the early Hawking radiation, one can distill a qubit that is perfectly entangled with an outgoing particle from the black hole~\cite{Almheiri13}. According to Hawking's semiclassical calculation~\cite{Hawking75}, however, that same particle is entangled with some field mode inside the black hole, leading to a violation of quantum mechanics, namely the monogamy of quantum entanglement. This contradiction, called the AMPS puzzle, sharply illustrates the fundamental difficulty in describing the interior of a quantum black hole.

The AMPS problem becomes most precise for the outside observer where the difficulty is constructing operators for interior modes. This viewpoint led to approaches bundled under ``$A=R_{B}$'' or ``$\text{ER} = \text{EPR}$'' where the interior operators are reconstructed in the early radiation instead of the remaining black hole~\cite{Almheiri13b, Papadodimas:2013aa, Maldacena13, Susskind13, Bousso:2013ab}. Unfortunately this proposal runs into various paradoxes~\cite{Marolf:2013aa, Bousso:2013aa, Giddings:2013aa}. (See~\cite{Almheiri13b, Harlow:2016ab} for summaries). The reconstruction may lead to apparent non-local interaction between the interior and distant radiation
~\footnote{The fact that some description of interior operator requires the early radiation does not necessarily mean that information can be sent non-locally. However, if every possible reconstruction of the interior operators requires the early radiation, then it is problematic as it implies that the infalling observer may interact non-locally with the early radiation by seeing interior modes. }. 
Relatedly, the interior operators seem to be sensitive to perturbations added to the early radiation. Also, previous proposals construct interior operators in a \emph{state-dependent} manner that suffers from a number of potential inconsistencies with quantum mechanics~\cite{Bousso:2014aa, Marolf:2016aa}. To overcome these difficulties, further proposals have been put forward. A concrete state-dependent proposal was made in the context of the AdS/CFT correspondence~\cite{Papadodimas:2013aa, Papadodimas:2014aa, Papadodimas:2014ab} and the Sachdev-Ye-Kitaev (SYK) model~\cite{Kourkoulou17, Almheiri2018}, but these proposals have some caveats~\cite{Harlow:2014aa}. 

Preliminary studies suggest that the idea of quantum error-correction~\cite{Almheiri:2015ac, Pastawski15b} plays key roles in avoiding some of the identified problems~\cite{Verlinde:2013aa, Almheiri2018, Verlinde13a, Verlinde13b, Goel18}. The ultimate resolution of the AMPS puzzle, however, would be obtained only by finding an explicit way of reconstructing interior operators which does not suffer from the aforementioned problems. 

Recently we pointed out that the procedure of reconstructing the interior operators can be viewed as the Hayden-Preskill recovery~\cite{Beni18}, a phenomenon of reconstructing an infalling quantum state from an old black hole by collecting the Hawking radiation~\cite{Hayden07}. Input quantum states can be recovered when the time-evolution of the system scrambles quantum information in a sense of decay of out-of-time ordered correlations~\cite{Hosur:2015ylk}. A quantum black hole exhibits such dynamics in the most dramatic manner~\cite{Maldacena:2016aa}. In parallel with the conclusion by Hayden and Preskill, this observation suggests that the interior operators can be constructed ``almost'' inside the remaining black hole with a few extra qubits from the early radiation. Also, the reason why interior operators are insensitive to perturbations on the early radiation can be explained by scrambling dynamics without relying on speculative quantum circuit complexity arguments~\cite{Harlow:2013aa, Maldacena13}. However, our observation in~\cite{Beni18} still suffers from the non-locality problem and state-dependence. 

In this paper, we expand on the aforementioned observation and present a protocol to reconstruct interior operators in a \emph{state-independent} way while at the same time avoiding the non-locality problem. In particular, we show that quantum entanglement between the outgoing mode and the reconstructed interior mode is insensitive to any perturbation on the early radiation. In fact, we argue that our construction of state-independent interior operators is applicable to generic black holes, young or old, in quasi thermal equilibrium. We will explain how our construction \emph{avoids} previous arguments which appear to suggest that interior operators must be state-dependent~\footnote{I thank Ahmed Almheiri and Suvrat Raju for useful discussions on this.}. 
Namely, we will find that construction of interior operators does not depend on the initial state of the black hole (\emph{i.e.} how it is entangled with the early radiation), but \emph{depend} on the infalling observer.

Based on the state-independent reconstruction, we suggest the following resolution of the AMPS puzzle. Let us distill a qubit $\tilde{D}$ from the early radiation $R$ that is entangled with the outgoing mode $D$. We show that an infalling observer leaves non-trivial gravitational backreaction which disentangles the outgoing mode $D$ from the distilled qubit $\tilde{D}$ \emph{no matter how she falls into a black hole}. The outgoing mode $D$ will be entangled with the interior mode $\overline{D}$ which is \emph{distinct} from the originally entangled qubit $\tilde{D}$. The infalling observer crosses the horizon smoothly and can observe entangled modes $D$ and $\overline{D}$. Namely, any quantum operation on the early radiation $R$ \emph{cannot} influence the experience of the infalling observer since the construction of the interior mode $\overline{D}$ does not involve any degrees of freedom on $R$. The disentangling phenomenon is a universal feature of a quantum black hole which can be studied quantitatively by using out-of-time order correlation functions. 

The aforementioned proposal provides an account for a physical mechanism where the infalling observer sees quantum entanglement while the outside observer does not. It is intriguing to consider an opposite limit of the monogamy relation where the outside observer can see quantum entanglement while the infalling observer does not.
Previous works, especially the one by Maldacena and Susskind~\cite{Maldacena13}, suggested that verification of quantum entanglement, namely the distillation of the entangled qubit $\tilde{D}$ on the early radiation $R$, generates a high-energy gravitational shockwave which would become a firewall and prevent the infalling observer from crossing the horizon smoothly. 
This proposal, however, can be refuted by observing that the interior mode $\overline{D}$ is distinct from the distilled qubit $\tilde{D}$ and its construction does not involve the early radiation $R$ at all. 
Here, we argue that the verification of quantum entanglement will generate ``safety barrier'' which stops the infalling observer from falling into a black hole. 
The key observation is that the outside observer can retrieve quantum entanglement back from the infalling observer by performing the Hayden-Preskill recovery on the infalling observer. Since the infalling observer does not cross the horizon, she will not be able to see quantum entanglement~\footnote{Here the infalling observer should be thought of as some light probe particle. If it were a macroscopic object with large energy, the outside observer wouldn't be able to save it by low quantum complexity operations. Also the amount of entanglement the outside observer needs to verify depends on the entropy associated with the infalling observer. See~\cite{Yoshida:2017aa} for details.}.
This proposal resonates with the recovery protocols for the Hayden-Preskill thought experiment proposed by the author and Kitaev~\cite{Yoshida:2017aa}. In fact, these recovery protocols proceed by protecting quantum entanglement between the outgoing mode $D$ and the early radiation $R$ from the backreaction by the infalling observer. 

While we rely on intuitions from the AdS/CFT correspondence, our arguments are applicable to generic quantum black holes. There are two main assumptions behind our proposals. First, the outside observer has a quantum description of a black hole with a pure quantum state on his Hilbert space. Second, out-of-time ordered correlation functions decay around/after the scrambling time. The proposed state-independent construction of interior operators work for generic quantum black holes that satisfy these two assumptions. 
Much of our argument, however, is based on simple quantum information theoretic description of a quantum black hole. While detailed calculations on realistic black holes are beyond the scope of the paper, we will try to be clear about our assumptions behind the use of simplified toy models~\footnote{Our attitude is to treat a black hole as a quantum system and try to explain the AMPS puzzle without giving up with unitarity, smoothness (equivalence principle) or locality.}.

The rest of the paper is organized as follows. 
In section~\ref{sec:AMPS}, we briefly review the firewall argument. 
In section~\ref{sec:HP}, we briefly review the Hayden-Preskill recovery and the quantum cloning puzzle. 
In section~\ref{sec:OTOC}, we discuss the notion of quantum information scrambling and its relation to out-of-time ordered correlation functions.
In section~\ref{sec:reconstruction}, we show that the interior operators can be reconstructed by using the Hayden-Preskill recovery protocol.
In section~\ref{sec:complexity}, we present a state-independent reconstruction of the interior operators.
In section~\ref{sec:outside}, we discuss the effect of backreaction by the infalling observer.
In section~\ref{sec:puzzle}, we propose the resolution of the AMPS puzzle.  
In section~\ref{sec:discussion}, we conclude with discussions. 
In appendix~\ref{sec:interaction}, we show that measurement of energy density suffices to entangle the infalling observer with the outgoing radiation. 
In appendix~\ref{sec:Verlinde}, we make comments on the proposal by Verlinde and Verlinde~\cite{Verlinde:2013aa}.
In appendix~\ref{sec:FAQ}, we discuss how our construction avoids previous arguments for the necessity of state-dependence. 
In appendix~\ref{sec:CMT}, we discuss the state-independent reconstruction from the perspective of dynamics in quantum many-body systems.
In appendix~\ref{sec:Alice}, we discuss geometric interpretations of the Hayden-Preskill recovery protocols.

\section{Review of firewall argument}\label{sec:AMPS}

We begin with a brief overview of the original argument of AMPS~\cite{Almheiri13}. The argument has a number of technical assumptions and fine details, and we will not cover all of them. Our goal is to remind readers of the form of quantum entanglement and the associated Hilbert space structure postulated by AMPS, as well as to introduce some notation. We also summarize some of the proposed resolutions of the puzzle, and potential difficulties and concerns that may arise from them. Our goal is to motivate some of the desired properties in reconstructing the interior operators. 

\begin{figure}
\centering
(a)\includegraphics[width=0.2\textwidth]{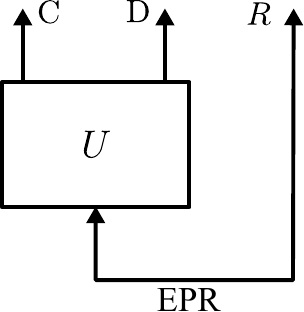} \qquad
(b)\includegraphics[width=0.29\textwidth]{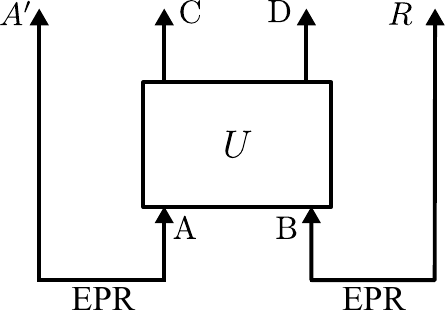} 
\caption{The notation used in the paper. Here $U$ represents the unitary time-evolution of the black hole. (a) The AMPS thought experiment. $C:$ the remaining black hole. $D:$ the outgoing mode. $R:$ the early radiation. (b) The Hayden-Preskill thought experiment. $A':$ the reference. $A$: the input quantum state. $B:$ the initial black hole. $R:$ the early radiation. $C:$ the remaining black hole. $D:$ the outgoing mode. Horizontal lines represent that two subsystems are in the EPR pairs: $\frac{1}{\sqrt{d}} \sum_{j=1}^{d} |j\rangle \otimes |j\rangle$.
}
\label{fig-notation}
\end{figure}

The notation in the present paper is slightly non-standard. For convenience, our notation is summarized in Fig.~\ref{fig-notation}. Throughout the paper, we will denote the infalling and outside observers by Alice and Bob respectively. 

\subsection{From the outside}\label{sec:AMPS_outside}

We first discuss the description of a quantum black hole from the perspective of Bob, the outside observer. 

\begin{figure}
\centering
(a)\includegraphics[width=0.2\textwidth]{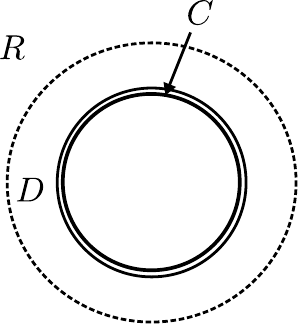} \qquad
(b)\includegraphics[width=0.2\textwidth]{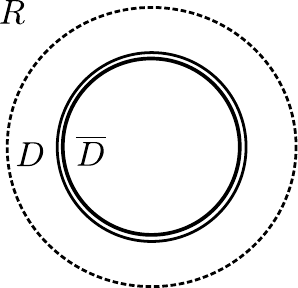} 
\caption{The Hilbert space structures. The dotted line represents $r=3GM$ and the double lines represent $r= 2GM \pm \epsilon$. Roughly, $D$ corresponds to the zone. (a) The outside Hilbert space $\mathcal{H}_{\text{outside}}$. (b) The infalling Hilbert space $\mathcal{H}_{\text{inside}}$.
}
\label{fig-Hilbert-space}
\end{figure}

The AMPS thought experiment relies on the following assumption on the Hilbert space structure. In Bob's description, the formation and evaporation of the black hole is described by a unitary operator. Namely, there is a pure quantum state $|\Psi\rangle$ in the outside Hilbert space $\mathcal{H}_{\text{outside}}$ at any given time. The Hilbert space $\mathcal{H}_{\text{outside}}$ can be factorized into (Fig.~\ref{fig-Hilbert-space}(a))
\begin{align}
\mathcal{H}_{\text{outside}} = \mathcal{H}_{C} \otimes \mathcal{H}_{D} \otimes \mathcal{H}_{R}.
\end{align}
Here $\mathcal{H}_{R}$ consists of the radiation field outside of the black hole with roughly $r > 3GM$ while $\mathcal{H}_{D}$ are the modes confined in $ 2GM + \epsilon < r < 3GM $. Here $r=2GM + \epsilon$ corresponds to the location of the stretched horizon, and $\epsilon$ is of order the Planck length. This region is often called the zone. It is common to restrict $\mathcal{H}_{D}$ to include only modes with Schwarzschild energy less than the black hole temperature since higher energy modes are not confined. Finally, $\mathcal{H}_{C}$ are the remaining degrees of freedom which can be interpreted as some entities sitting at the stretched horizon at $r = 2GM + \epsilon$. It is common to restrict $\mathcal{H}_{R}$ to be a finite dimensional Hilbert space for convenience of discussions. Precise distinctions among these subsystems are not essential for the AMPS puzzle.

Following AMPS, we consider an old black hole that has already emitted more than half of its initial entropy: $|R| \gg |C||D|$. If $|\Psi\rangle$ is a Haar random quantum state in $\mathcal{H}_{\text{outside}}$~\cite{Page93, Page:1993aa}, the density operator in $CD$ is approximately proportional to the identity operator:
\begin{align}
\rho_{CD} \approx \frac{1}{|C||D|} I_{C} \otimes I_{D}. \label{eq:no_correlation}
\end{align}
This implies $D$ is maximally entangled with $R$. Of course, approximating $|\Psi\rangle$ with a Haar random state is an oversimplification. A more careful analysis suggests that thermal correlations are present in $\rho_{CD}$~\cite{Verlinde:2013aa} although these are purely classical ones resulting from the conservation of the energy~\cite{Beni18}. For the time being, we proceed with Eq.~\eqref{eq:no_correlation}. We emphasize that this detail does not matter for discussions except appendix~\ref{sec:interaction}. 

\subsection{From the inside}

Next we discuss the description from the perspective of Alice, the infalling observer.

Following AMPS, we assume that Alice has quantum mechanical description of her experiences in terms of a quantum state on her time slice. Namely, the Hilbert space $\mathcal{H}_{\text{outside}}$ can be factorized into (Fig.~\ref{fig-Hilbert-space}(b))
\begin{align}
\mathcal{H}_{\text{inside}} = \mathcal{H}_{\overline{D}} \otimes \mathcal{H}_{D} \otimes \mathcal{H}_{R} \otimes \mathcal{H}_{S}.
\end{align}
Here $\mathcal{H}_{D}$ and $\mathcal{H}_{R}$ correspond to the confined modes (the zone) and the radiation fields respectively as defined above. These are subsystems which both Alice and Bob can access. $\mathcal{H}_{\overline{D}}$ are the modes just inside the horizon roughly with $GM < r < 2GM -\epsilon$. Finally, $\mathcal{H}_{S}$ is all the remaining degrees of freedom that is not accessible to Alice. The modes at the stretched horizon $\mathcal{H}_{C}$ are often included in $\mathcal{H}_{S}$, but this detail does not matter in the AMPS argument. Finally, we postulate that the quantum description of $\rho_{DR}$ for Alice and Bob are identical~\footnote{By ``identical'', we mean that $\rho^{(\text{Alice})}_{DR}=\rho^{(\text{Bob})}_{DR}$. It may be possible that $\rho^{(\text{Alice})}_{DR}$ and $\rho^{(\text{Bob})}_{DR}$ differ up to local unitary transformations with factorized form $U_{D}\otimes U_{R}$. The key assumption for us is that non-local properties such as quantum entanglement between $D$ and $R$ are identical for two observers.}. 

The AMPS puzzle concerns an apparent inconsistency between Alice and Bob's descriptions. In Alice's description, $\mathcal{H}_{\overline{D}}$ and $\mathcal{H}_{D}$ contain Rindler modes which are close to maximally entangled:
\begin{align}
\propto \sum_{n}  e^{- \frac{\beta E_{n}}{2}} |n\rangle_{\overline{D}} \otimes |n\rangle_{D}. 
\end{align}
Alice's Hilbert space may contain some degrees of freedom other than Rindler modes. The important point for us is that two modes in $\mathcal{H}_{\overline{D}}$ and $\mathcal{H}_{D}$ must be entangled. On the other hand, in Bob's description, $D$ is maximally entangled with the radiation $R$, which would imply that $D$ and $\overline{D}$ are not entangled with each other due to the monogamy of entanglement. In particular, by using the strong subadditivity, AMPS, as well as earlier works~\cite{Mathur:2009aa, Braunstein:2013aa}, arrived at $I(D,\overline{D})\equiv S_{D} + S_{\overline{D}} - S_{D\overline{D}} \approx 0$. This suggests $\rho_{\overline{D}D} = \rho_{\overline{D}}\otimes \rho_{D}$, the existence of a firewall~\footnote{More precisely, this suggests a strong breakdown of the equivalence principle (or effective quantum field theory description) in a regime of low curvature.}~\footnote{The term ``firewalls'' generically denote various forms of high energy densities that may appear near the horizon. In addressing the AMPS puzzle, we will focus on firewalls that may arise from the loss of quantum entanglement between pairs of modes inside and outside the black hole. In other words, our primary focus is on an apparent violation of monogamy of quantum entanglement and how to restore it by properly understanding the breakdown of semiclassical arguments through the lens of quantum information theory. Note that there may be other physical mechanisms for firewalls, see~\cite{Mathur:2009aa} for instance, but these are beyond the scope of this paper.} 
. 

\subsection{Interior operators}

The AMPS puzzle can be also stated in the outside observer's description. This viewpoint becomes most precise in the AdS/CFT correspondence where $R$ is interpreted as degrees of freedom on the other side of a large two-sided AdS black hole. In Bob's description, or in the boundary CFT, the puzzle can be identified with the fact that $I(C,D)=0$. This suggests that the interior mode in $\overline{D}$ cannot be described as an operator inside $\mathcal{H}_{C}$. Namely, its description must involve the radiation $R$. This viewpoint is often advocated under the slogan ``$\overline{D} = R$'' (or ``$A = R_{B}$'' in accordance with the standard notation in the literature) and also ``$\text{ER}=\text{EPR}$''. This proposal, however, runs into various paradoxes. 

First, it may lead to non-local signalling from the interior of the black hole to the radiation $R$ (and vice versa). To be concrete, one may consider a large AdS black hole with very long wormhole throat. The description of the interior mode, which an infalling observer would be able to measure, \emph{always} requires degrees of freedom from a far distant boundary of the black hole, which sounds highly unphysical. The non-local signalling problem is referred to as the commutator problem since operators for the interior mode $\overline{D}$ do not commute with operators in the radiation $R$. To avoid the non-local signalling problem, the reconstruction of the interior operator in the outside should be insensitive to perturbations on $R$. The problem may be resolved by employing the idea of quantum error-correction where the interior operator may be encoded in $R$ via some sort of quantum error-correcting code~\cite{Almheiri:2015ac, Pastawski15b}. This viewpoint has been considered seriously by several authors~\cite{Verlinde:2013aa, Almheiri2018, Verlinde13a, Verlinde13b, Goel18}. Relatedly, arguments based on quantum circuit complexity have been attempted~\cite{Harlow:2013aa, Maldacena13}. The key question, however, is the physical origin of fault-tolerant encoding of the interior operators~\footnote{Some previous works argue that the involvement of the early radiation $R$ is fine as long as information cannot be sent non-locally. We think that this is a serious problem if all the possible representations of the interior mode require degrees of freedom from $R$ since the measurement of the interior mode by the infalling observer would require non-local coupling between the remaining black hole $C$ and the early radiation $R$. In this paper, we will show that the construction of the interior mode and its measurement by the infalling observer can be achieved without ever involving degrees of freedom on the early radiation $R$.}.

Second, it leads to state-dependent construction of interior operators~\cite{Papadodimas:2013aa, Papadodimas:2014aa, Papadodimas:2014ab}. To see the origin of state-dependence, suppose that a quantum state of a black hole $|\Psi\rangle$ is given by some maximally entangled state with $S_{CD}=\log_{2} |C||D|$. Such a state can be written as 
\begin{align}
|\Psi\rangle = (I \otimes K) |\Phi^{\text{EPR}}\rangle, \qquad |\Phi^{\text{EPR}}\rangle_{CDR} \equiv \frac{1}{\sqrt{|C||D|}}\sum_{i,j}|i,j\rangle_{CD} \otimes |i,j\rangle_{R} \label{eq:complex_EPR}
\end{align}
where $K$ is some unitary operator and $|\Phi^{\text{EPR}}\rangle_{CDR}$ are EPR pairs between $CD$ and $R$
~\footnote{
Harlow and Hayden suggested that there is a certain obstruction for the infalling observer to verify the quantum entanglement between the outgoing mode $D$ and the radiation $R$, hence the AMPS thought experiment cannot be carried out~\cite{Harlow:2013aa}. Namely, they conjectured that quantum circuit complexity of verifying the quantum entanglement is exponential in the entropy $S_{\text{BH}}$ by relying on computational complexity theoretical assumptions. This viewpoint has been further strengthened by Aaronson and Susskind~\cite{Susskind18, Aaronson16}. However, the complexity theoretic obstruction itself can be avoided by performing the so-called precomputation~\cite{Oppenheim:2014aa}. Also, the argument does not apply to the eternal AdS black hole where the verification of entanglement can be performed via simple quantum operations.
}. 
Then construction of operators for the interior modes $\overline{D}$ on the early radiation $R$ depends on the unitary operator $K$. But these state-dependent interior operators are problematic in describing quantum physics of the infalling observer due to a number of potential inconsistencies with quantum mechanics~\cite{Marolf:2013aa, Bousso:2014aa, Bousso:2013aa, Bousso:2013ab, Marolf:2016aa, Harlow:2014aa}~\footnote{Again, we think that this is a serious problem if all the possible representations of the interior mode are state-dependent.}. 

To summarize, our goal is to reconstruct interior operators in a state-independent manner with fault-tolerance against perturbations on $R$ and without the non-locality problem. 

\section{Review of Hayden-Preskill recovery}\label{sec:HP}

In this section, we provide a brief review of the Hayden-Preskill thought experiment~\cite{Hayden07}. We try to be clear about what the recoverability of quantum states means in information-theoretic terms. 
We will mostly focus on a situation which mimics the AdS eternal black hole while the scope of the original argument by Hayden and Preskill is broader. 
Part of our goal is to remind readers of how the quantum cloning puzzle can be resolved by using the idea of backreaction within the context of the AdS/CFT correspondence. 

\begin{figure}
\centering
\includegraphics[width=0.22\textwidth]{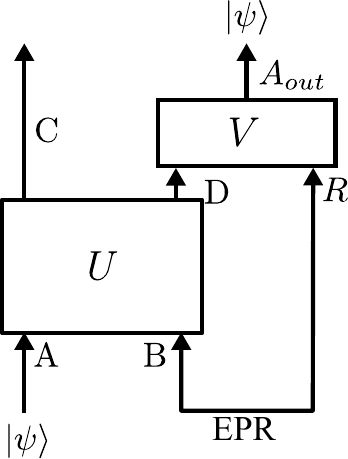} 
\caption{The Hayden-Preskill thought experiment. $U$ represents the time-evolution of the black hole and $V$ represents a recovery unitary.
}
\label{fig-HP}
\end{figure}

\subsection{Setup}

As in the firewall argument, Hayden and Preskill considered an old black hole which has emitted more than half of its initial entropy. In the Hayden-Preskill thought experiment, Alice throws a quantum state $|\psi\rangle$ into a black hole and Bob, the outside observer, attempts to reconstruct it by collecting the Hawking radiation. 
To simplify the argument further, we treat the initial state as $n_{B} = \log |B|$ copies of EPR pairs between the black hole $B$ and the radiation $R$
:
\begin{align}
|\Phi^{\text{EPR}} \rangle_{BR} = \frac{1}{\sqrt{|B|}}\sum_{j} |j\rangle_{B}\otimes |j\rangle_{R}. \label{eq:EPR_initial}
\end{align}

Let us append a subsystem $A$ to the black hole to account for the Hilbert space of the infalling quantum state $|\psi\rangle$. Following Hayden and Preskill, let us assume that the black hole evolves by a Haar random unitary $U$ that acts on $AB$. Let $C$ and $D$ be the remaining black hole and the outgoing mode respectively (Fig.~\ref{fig-HP}). Bob's goal is to reconstruct $|\psi\rangle$ by catching $D$ \emph{and} having access to the early radiation $R$. The surprising result is that $|D|\gtrapprox |A|$ is sufficient to achieve this goal. Namely, if $|\psi\rangle$ is an $n_{A}$-qubit quantum state, catching $n_{D} = n_{A} +\epsilon$ qubits of the outgoing mode with $\epsilon = O(1)$ suffices. By ``reconstruction'', we mean the existence of some recovery unitary $V_{DR}$ that, for any given input state $|\psi\rangle$, reconstructs the original state $|\psi\rangle$ (Fig.~\ref{fig-HP}):
\begin{align}
(I_{C}\otimes V_{DR} )(U_{AB}\otimes I_{R})\big( |\psi\rangle_{A} \otimes |\Phi^{\text{EPR}} \rangle_{BR}\big) \approx |\psi\rangle_{A_{\text{out}}} \otimes |\text{something}\rangle \qquad \text{for all} \quad |\psi\rangle \in \mathcal{H}_{A}.
\end{align}
Here $A_{\text{out}}$ is some subsystem in $DR$ with $|A_{\text{out}}|=|A|$,  and ``$\approx$'' is measured in terms of the fidelity of the output state on $A_{\text{out}}$ with $|\psi\rangle$ (i.e., the expectation value of the output state with respect to the projector $|\psi\rangle\langle\psi|_{A_{\text{out}}}$).

\subsection{State representation of $U$}

Instead of throwing an unknown quantum state $|\psi\rangle$ into a black hole, it is more convenient to introduce an ancillary subsystem $A'$ which has the same dimensionality as $A$. We will see that this trick enables us to discuss the recoverability in a quantitative manner with information-theoretic measures. Prepare an EPR pair on $A'$ and $A$ and throw qubits from $A$ into the black hole $B$. After the time-evolution by a unitary operator $U$ that acts on $AB$, the system is given by
\begin{align}
|\Psi\rangle = \big(I_{A'} \otimes U_{AB} \otimes I_{R}\big) \big(|\text{EPR}\rangle_{A'A}\otimes |\Phi^{\text{EPR}}\rangle_{BR} \big) = \figbox{1.0}{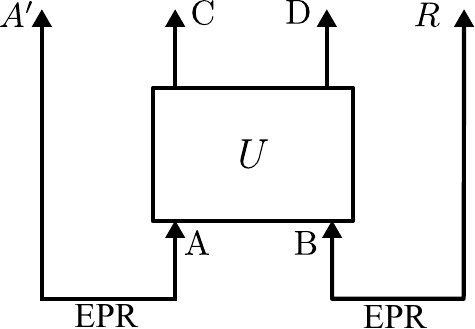}
\label{eq:state}
\end{align} 
which is supported on $A'CDR$. The wavefunction in Eq.~\eqref{eq:state} is often called the ``state representation'' (or the ``Choi representation'') of $U$~\cite{Hosur:2015ylk}. To remind readers of the notation, $B$ is the initial black hole, $R$ is the early radiation, $C$ is the remaining black hole and $D$ is the late radiation.

Information theoretically, the recoverability of unknown quantum states can be quantitatively addressed by the amount of entanglement between $A'$ and $RD$, e.g., the mutual information $I(A',DR)$. If $I(A',DR)$ is close to its maximal value, nearly perfect EPR pair can be distilled by applying some recovery unitary operator $V_{DR}$:
\begin{align}
(I_{A'C}\otimes V_{DR})|\Psi\rangle \approx |\text{EPR}\rangle_{A'A_{\text{out}}} \otimes |\text{something}\rangle.
\end{align}
Here ``$\approx$'' is measured in terms of the fidelity with the perfect EPR pair (i.e., the expectation value with respect to the EPR projector on $A'A_{\text{out}}$)~\footnote{The fidelity of the distilled EPR pair can be quantitatively related to out-of-time ordered correlation function~\cite{Yoshida:2017aa}.}.

Distillation of EPR pair can be related to recovery of infalling quantum states as follows. By projecting the reference system $A'$ onto $|\psi^{*}\rangle$, we will have $|\psi\rangle$ on the original input Hilbert space $A$ since
\begin{align}
(\langle  \psi^{*}|_{A'} \otimes I_{A} ) |\text{EPR}\rangle_{A'A} \propto |\psi\rangle_{A}.
\end{align}
 If $|\text{EPR}\rangle_{A'A_{\text{out}}}$ is distilled on $A'A_{\text{out}}$, the above projection on $A'$ will generate $|\psi\rangle_{A_{\text{out}}}$ on the output Hilbert space $A_{\text{out}}$, implying successful reconstruction of the input state. One merit of introducing the reference system $A'$ is that we do not need to keep track of how the recovery operation works for each choice of input states.

\subsection{Reconstruction of operators}

We will use the Hayden-Preskill recovery protocols to construct interior operators in the AMPS problem. For this purpose, it is convenient to state the Hayden-Preskill recovery in terms of operators. Given an arbitrary unitary operator $O_{A'}$ acting on $A'$, it is possible to identify a partner operator on $DR$ if $I(A',DR)$ is nearly maximal (i.e., OTOCs are small). Recall the following relation:
\begin{align}
(O_{A'} \otimes I )|\text{EPR}\rangle_{A'A_{\text{out}}} = (I \otimes O_{A_{\text{out}}}^{T} )|\text{EPR}\rangle_{A'A_{\text{out}}}
\end{align}
or graphically 
\begin{align}
\figbox{1.0}{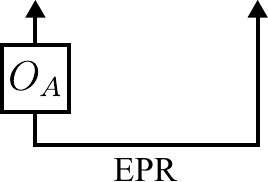}\ = \ \figbox{1.0}{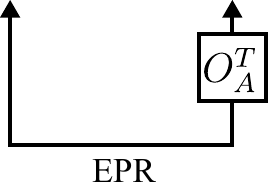}
\end{align}
which holds for any unitary operator $O_{A'}$ and its transpose $O_{A_{\text{out}}}^{T}$. Define $\widetilde{O}_{DR}^T$  as follows:
\begin{align}
\widetilde{O}_{DR}^T \equiv V^{\dagger}_{DR} O_{A_{\text{out}}}^T  V_{DR}.
\end{align}
Then, the partner operator $\widetilde{O}_{DR}^{T}$ on $DR$ satisfies 
\begin{align}
(O_{A'} \otimes I_{CDR})|\Psi\rangle \approx (I_{A'C} \otimes \widetilde{O}_{DR}^{T})|\Psi\rangle
\end{align}
or graphically 
\begin{align}
\figbox{1.0}{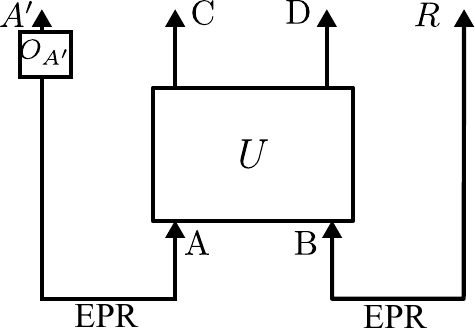}\ \approx \ \figbox{1.0}{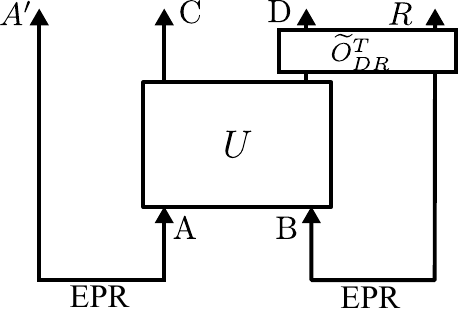}. 
\end{align}

\subsection{Quantum cloning puzzle}

The Hayden-Preskill recovery leads to an apparent cloning of quantum states since Alice possesses $|\psi\rangle$ inside the black hole whereas Bob possesses the reconstructed copy of $|\psi\rangle$ outside the black hole~\footnote{By introducing the reference system $A'$, the quantum cloning puzzle can be stated as potential violation of monogamy of entanglement. This observation hints that the Hayden-Preskill thought experiment is indeed related to the firewall problem.
}. A traditional approach to resolve this puzzle is to resort to the idea of complementarity; we decide not to be bothered by violation of quantum mechanics as long as there is no observer who can verify it~\cite{t-Hooft:1990aa, t-Hooft:1985aa, Susskind:1994aa, Susskind:1993aa}. In order to verify the quantum cloning, Bob needs to wait for a black hole to implement a highly-complicated unitary $U$, catch the outgoing mode, reconstruct $|\psi\rangle$, and jump into a black hole to meet Alice who possesses the other copy. For the time being, assume that reconstructing $|\psi\rangle$ is instantaneous once the outgoing mode is collected. Then if it takes more than $t \sim r_{s} \log r_{s}$ in the Schwarzschild time (or $t\sim \log r_{s}$ at the stretched horizon) for the black hole to implement $U$, Alice will reach the black hole singularity before meeting Bob. Quantum information theoretic studies suggest that time scale to ``delocalize'' quantum information must be indeed longer than $t\sim \log r_{s}$ in a clock at the stretched horizon~\cite{Lashkari13}. Hence, the quantum cloning puzzle is avoided. 

While the above viewpoint provides a tentative resolution of the quantum cloning puzzle, a refined resolution has been recently proposed within the context of the AdS/CFT correspondence. The original argument via complementarity assumes that the black hole horizon is an absolute entity and its location never moves regardless of quantum operations applied from the outside. A modern viewpoint on this puzzle is to postulate that Bob's recovery operation may have a non-trivial backreaction to the black hole geometry. This viewpoint is largely motivated from recent developments in studies of backreactions in the AdS/CFT correspondence and related toy models such as the SYK model~\cite{Kitaev_unpublished, Almheiri:2015ab, Maldacena:2016ab, Engelsoy:2016aa, Stanford:2016aa}. The key insight is that Bob's operation from the outside may shift the location of the event horizon and pull the quantum state $|\psi\rangle$ from the interior to the exterior. A concrete physical realization of this scenario is the so-called traversable wormhole phenomena discovered by Gao, Jafferis and Wall~\cite{Gao:2017aa}. In particular, for the eternal AdS black hole, they identified certain forms of interactions between CFTs on opposite boundaries which send negative energy to the bulk and shift the event horizon so that signals can traverse the wormhole~\footnote{Since their interaction couples two boundaries, null geodesics through the wormhole is no longer achronal. Hence the average null energy condition (ANEC) is obeyed for achronal geodesics.}. The relation between traversable wormholes and the Hayden-Preskill thought experiment is also discussed in~\cite{Traversable2017, Yoshida:2017aa}.

\section{Scrambling and recovery}\label{sec:OTOC}

We have reviewed the Hayden-Preskill thought experiment and the quantum cloning puzzle from the perspective of backreaction within the context of the AdS/CFT correspondence, namely for the eternal AdS black hole. In this section, we extend the argument to more generic quantum systems by reviewing the relation between the Hayden-Preskill recovery and quantum information scrambling as diagnosed by out-of-time ordered correlation (OTOC) functions. 
We also discuss concrete recovery protocols for the Hayden-Preskill thought experiment.

\subsection{Scrambling}

Let us continue our discussion on the Hayden-Preskill recovery. Again, we will focus on a quantum black hole whose initial state is represented by 
$|\Phi^{\text{EPR}} \rangle_{BR} = \frac{1}{\sqrt{|B|}}\sum_{j} |j\rangle_{B}\otimes |j\rangle_{R}$.
We are primarily interested in quantum entanglement between $A'$ (the reference system) and $DR$ (the late radiation and the early radiation) in the state representation of $U$ as defined in Eq.~\eqref{eq:state}. 

While Hayden and Preskill considered Haar random unitary $U$, a recovery operation can be performed in strongly-interacting quantum systems which delocalize quantum information over the whole system. This phenomena, often called quantum information scrambling, can be probed by the out-of-time ordered correlation (OTOC) function~\cite{Larkin68, Kitaev_unpublished, shenker2014black, Roberts:2015aa}:
\begin{align}
\langle O_{A}(0) O_{D}(t) O_{A}^{\dagger}(0) O_{D}^{\dagger}(t) \rangle \equiv \frac{1}{d}\Tr\big( O_{A} U^{\dagger}O_{D}U  O_{A}^{\dagger} U^{\dagger}O_{D}^{\dagger} U  \big)
\end{align}
where $d= |A||B|=|C||D|$ denotes the total Hilbert space dimension. Two different bipartitions of the total Hilbert space $\mathcal{H}$, into $\mathcal{H}_{A} \otimes \mathcal{H}_{B}$ and $\mathcal{H}_{C} \otimes \mathcal{H}_{D}$, are considered. Here $U$ is an arbitrary unitary operator that accounts the time-evolution of the system. Note that $O_{A}$ and $O_{D}$ are operators that act on subsystems $A$ and $D$ respectively. For concreteness, we take $O_{A},O_{D}$ to be basis operators (such as Pauli operators) supported on $A,D$. For $A,D$ with no overlap, the OTOC starts at $1$ and then decays under chaotic time-evolution. 

\begin{figure}
\centering
\includegraphics[width=0.15\textwidth]{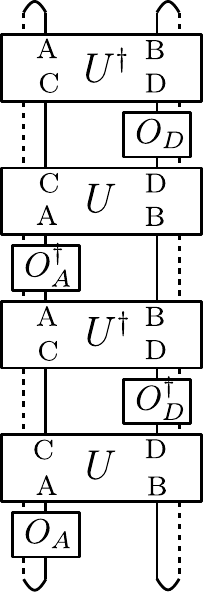} 
\caption{Out-of-time ordered correlation (OTOC) functions. 
}
\label{fig-clarification}
\end{figure}

The decay of OTOCs implies that the Hayden-Preskill recovery can be performed~\cite{Hosur:2015ylk, Roberts:2017aa}. Specifically, the following equality was proven:
\begin{align}
2^{- I^{(2)}(A',RD)  }  = \int d O_{A} d O_{D}\ \langle O_{A}(0) O_{D}(t) O_{A}^{\dagger}(0) O_{D}^{\dagger}(t) \rangle  \label{eq:OTOC_recovery}
\end{align}
where integrals with $d O_{A}, d O_{D}$ take averages over all the basis operators supported on $A,D$ respectively. Here $I^{(2)}(A',RD) \equiv S_{A'}^{(2)} + S_{RD}^{(2)} - S_{A'RD}^{(2)} $ represents the R\'{e}nyi-$2$ mutual information~\footnote{Whilt we have $I(A',RD) \geq I^{(2)}(A',RD)$ in the above setting, the R\'{e}nyi-$2$ mutual information does not lower bound the ordinary mutual information in general. In fact, from the perspective of quantum information theory, the R\'{e}nyi-$2$ mutual information is not a proper entanglement monotone. Instead one must use the Sandwiched R\'{e}nyi-$\alpha$ divergence. A similar duality relation is known for two of them with $\alpha$ and $\beta$ satisfying $\frac{1}{\alpha} + \frac{1}{\beta} = 2$~\cite{Beigi:2013aa}. In~\cite{Yoshida:2019aa}, a certain average of OTOCs is expressed as the $\alpha=2$ Sandwiched R\'{e}nyi divergence and is proven to lower bound the mutual information $I(A',RD)$. }.
It is worth reminding that the left hand side of Eq.~\eqref{eq:OTOC_recovery} is defined for the state representation of $U$ (Eq.~\eqref{eq:state}) which is supported on $\mathcal{H}_{A'}\otimes \mathcal{H}_{C}\otimes \mathcal{H}_{D}\otimes \mathcal{H}_{R} $ whereas OTOCs on the right hand side are calculated on $\mathcal{H}=\mathcal{H}_{A} \otimes \mathcal{H}_{B}=\mathcal{H}_{C} \otimes \mathcal{H}_{D}$. See Fig.~\ref{fig-clarification} for graphical representation of OTOCs. We emphasize that the aforementioned relation holds for any unitary operator $U$.

The Hayden-Preskill recovery and quantum information scrambling can be explicitly related by the above formula in Eq.~\eqref{eq:OTOC_recovery}~\footnote{An alternative definition of quantum information scrambling was previously considered~\cite{Sekino08, Lashkari13}. This definition considers initial quantum states with low entanglement (e.g., product states) and asks if the states become thermalized. We think that the definition based on OTOCs is a more appropriate characterization of quantum information scrambling as its relation to black hole physics and Hayden-Preskill thought experiment is clear. For interested readers, a definition of scrambling at finite temperature was proposed in~\cite{Yoshida:2017aa} by using OTOCs. Recently Shor presented an argument showing that scrambling in the alternative definition cannot be achieved within the scrambling time $\approx r_{s}\log r_{s}$~\cite{Shor18}.} . 
Namely, smallness of OTOCs implies large $I^{(2)}(A',RD) $ which implies the existence of reconstruction procedures
~\footnote{Note that the above formula is restricted to quantum systems at infinite temperature (i.e., the system is finite-dimensional and the quantum state in OTOCs is given by the maximally mixed state $\rho=\frac{1}{d}I$). As such, the conclusion on the recoverability is applicable only to quantum black holes whose initial states are approximated by maximally entangled states $\approx (I \otimes K )|\Phi^{\text{EPR}}\rangle_{BR}$. Essentially, these states are simplified descriptions of maximally entangled black holes as an $S_{\text{BH}}$-qubit system where $S_{\text{BH}}$ is the Bekenstein-Hawking (or the coarse-grained) entropy. 
Some readers might be worried about the validity of the relation between OTOCs and the recoverability for more realistic black holes at finite temperature. 
Further generalization of this relation for black holes with more generic quantum states has been discussed in~\cite{Yoshida:2017aa}. Namely it was shown that the decay of OTOCs implies the recoverability as long as quantum correlations between $A$ and $B$ as well as $C$ and $D$ are small.
For this reason, we do not view the simplification with $\approx (I \otimes K )|\Phi^{\text{EPR}}\rangle_{BR}$ as fundamental limitation of our argument. 
For simplicity of discussion, we proceed with this simplified description of a quantum black hole as a $S_{\text{BH}}$-qubit system. We will discuss its validity from other perspective in section~\ref{sec:code}. 
}.

\subsection{OTOCs in a black hole}

Having seen that scrambling implies the recoverability in the Hayden-Preskill thought experiment, let us discuss OTOCs for black holes. OTOCs for gravitational systems can be explicitly computed by taking $O_{A}$ to be a weak perturbation. Under the gravitational blueshift, such a perturbation is greatly amplified and becomes a gravitational shockwave as it falls into a black hole. For the Schwarzschild black hole, the effect of gravitational shockwaves on correlation functions were discussed by 't Hooft and Dray~\cite{Dray:1985aa, Hooft:1987aa}, and also by Kiem, Verlinde and Verlinde~\cite{Kiem:1995aa}. Kitaev pointed out that these pioneering works actually computed OTOCs in disguise~\cite{Kitaev_unpublished}. For the AdS black hole, OTOCs have been computed by Shenker and Stanford~\cite{shenker2014black}. Shockwave geometries in the Reissner-Nordstr\"{o}m black hole were studied by Sfetsos~\cite{Sfetsos:1995aa}. Recently shockwave geometries in the Kerr black hole were studied by BenTov and Swearngin~\cite{BenTov:2019aa}. 

These calculations suggest that OTOCs start to decay from its initial value significantly around the scrambling time, which is of order $\approx r_{s}\log r_{s}$ in the Schwartzshild black hole. As such, after the scrambling time, a quantum state that has fallen into a black hole can be retrieved as long as the black hole is maximally entangled. We emphasize that this argument is not restricted to black holes that are described in the AdS/CFT correspondence. Namely, we only assumed that the outside observer has a quantum description of a black hole with a pure quantum state on his Hilbert space, and OTOCs decay after the scrambling time.

\subsection{Reovery protocols}\label{sec:protocol}

The above relation in Eq.~\eqref{eq:OTOC_recovery} provides a formal proof of the existence of recovery protocols in a scrambling system. The author and Kitaev proposed two particular ``simple'' recovery protocols which work for any scrambling systems~\cite{Yoshida:2017aa}. Here we briefly review these protocols~\footnote{
These recovery protocols work well after the scrambling time whereas the traversable wormhole picture is valid before the scrambling time. The nature of scrambling after the scrambling time is quite different from that before the scrambling time, see~\cite{Kitaev:2018aa, Gu18} for instance. While this paper focuses on the physics after the scrambling time, we believe that a qualitatively similar implication on the AMPS puzzle can be made by relying on the physics of scrambling before the scrambling time. 
}.

\begin{figure}
\centering
\includegraphics[width=0.4\textwidth]{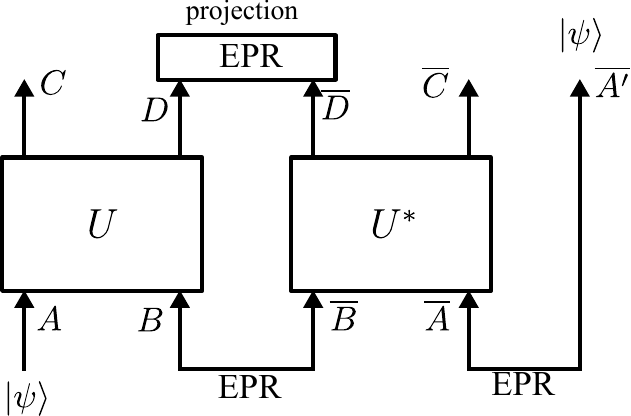} 
\caption{A recovery protocol for the Hayden-Preskill thought experiment from~\cite{Yoshida:2017aa}. The input quantum state $|\psi\rangle$ on $A$ is teleported to $\overline{A'}$ upon postselecting $D\overline{D}$ to be EPR pairs. In the diagram, $|A|=|\overline{A}| = |\overline{A'}|$, $|B|=|\overline{B}|$, $|C|=|\overline{C}|$, and $|D|=|\overline{D}|$. Here $D\overline{D}$ are interpreted as outgoing modes. For scrambling unitary $U$, the recovery protocol works when $|D|\gtrapprox |A|$. In particular, $D$ can be chosen as any subsystem as long as OTOCs $\langle O_{A}(0) O_{D}(t) O_{A}(0) O_{D}(t) \rangle$ are small.
}
\label{fig-HP-decode}
\end{figure}

The first protocol works probabilistically. It applies the complex conjugate $U^{*}$ to the entangled partner $R$ (which is denoted as $\overline{B}$ in Fig.~\ref{fig-HP-decode}), and then projects the outgoing modes $D$ and $\overline{D}$ from both sides into EPR pairs. Upon postselecting the measurement outcome to be EPR pairs, Alice's quantum state $|\psi\rangle$ will be teleported to Bob's register qubits on $\overline{A'}$. See Fig.~\ref{fig-HP-decode} for summary of the whole process. Noting that $D\overline{D}$ were initially EPR pairs at $t=0$ (if $A$ and $D$ do not overlap), this recovery operation can be interpreted as the outside observer's attempt to restore $D\overline{D}$ to be original EPR pairs. The protocol achieves high recovery fidelity when OTOCs decay.

The second protocol is more involved and works deterministically. It incorporates the quantum Grover search algorithm where $U^{*}$ and $U^{T}$ are applied on the entangled partner multiple times with Grover rotation operators inserted. Interested readers should see the original paper for details. Unlike the first protocol, the second protocol applies a careful sequence of unitary operators which restores $D\overline{D}$ to be EPR pair in a deterministic manner. If $|A|\sim O(1)$, the quantum circuit complexity of both recovery protocols is roughly equal to that of $U$. Hence, if it takes time $t$ to implement $U$, it takes $\approx t$ to reconstruct $|\psi\rangle$. Both protocols enable us to reconstruct partner operators, although the first protocol requires us to normalize reconstructed operators properly
~\footnote{
Some readers may be worried about the fact that Bob needs to understand the dynamics of quantum black hole for reconstruction. Given our little understanding of quantum gravity, this might sound like a valid concern. However we emphasize that this is not the issue intrinsically related to quantum gravity. In fact, a similar problem can be identified in classical systems. Suppose that Alice prepares a classical bit $0$ or $1$ and Bob tries to find out what is written in Alice's bit. If Alice's bit remains unchanged, Bob will immediately know the answer. If Alice has applied a Pauli-$X$ operator, her bit is flipped; $0 \rightarrow 1$ and $1 \rightarrow 0$. Then, Bob needs to flip her bit back to find the original information. Now suppose Bob does not know if Alice has applied $X$ or not. Then it is not possible for Bob to know Alice's original bit. As this simple classical observation suggests, Bob knowing the dynamics $U$ is a prerequisite to treat the information loss problem in a rigorous information theoretic setting, and our little knowledge of $U$ is a separate problem.
}
. 
Geometric interpretations of these protocols are discussed in Appendix~\ref{sec:Alice}. 

So far we have mostly focused on a quantum black hole whose quantum state can be approximately modeled by $\approx |\Phi^{\text{EPR}} \rangle_{BR}$. A generic quantum state for a maximally entangled black hole, however, should be modeled by $\approx (I \otimes K)|\Phi^{\text{EPR}} \rangle_{BR}$ as in Eq.~\eqref{eq:complex_EPR} with some unitary operator $K$. Here we would like to clarify the applicability of our argument. 

As a matter of principle, the quantitative argument on the recoverability remains valid regardless of $K$ since $K$ does not affect the mutual information $I(A',DR)$. In particular, as long as the unitary time-evolution operator $U$ forces OTOCs to decay, one can prove the existence of recovery protocols. Also, the aforementioned recovery protocols can be applied to generic initial states by replacing $U$ with $UK^{*}$. This essentially cancels the unitary operator $K$ and reduces the system back to $|\Phi^{\text{EPR}} \rangle_{BR}$. Performing recovery protocols with $UK^{*}$, however, may be problematic if $K$ is a highly complex unitary operator. One might argue that recovery protocols cannot be performed before the black hole evaporates completely. The difficulties associated with highly complex $K$ may be avoided by the idea of the ``precomputation''~\cite{Oppenheim:2014aa}. Yet, it is not clear if such a procedure is physically plausible in realistic systems.

We will return to this issue of the $K$-dependence in Section~\ref{sec:complexity} when the Hayden-Preskill recovery is used to reconstruct interior operators. We will see that, in the context of the AMPS puzzle, we can actually avoid the $K$-dependence and achieve state-independent reconstruction of interior operators. 

\section{Reconstruction of interior operators}\label{sec:reconstruction}

In this section, we point out that the black hole interior operators can be reconstructed by using a protocol similar to the Hayden-Preskill recovery procedure. This subsection is a slightly reorganized reprint from Section 5 of our previous work~\cite{Beni18}. We will focus on cases where the initial state of the old black hole is given by EPR pairs while deferring discussions on generic entangled states, as well as the issue of the state-dependence, to the next section. 

\subsection{Hayden-Preskill for interior operators}

The quantum state of interest is as follows:
\begin{align}
|\Psi\rangle = \figbox{1.0}{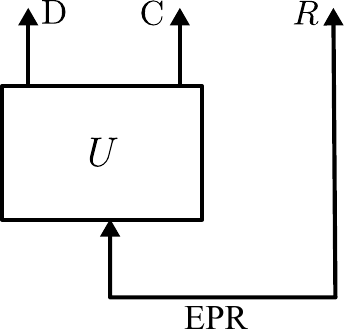} \label{eq:old-state}
\end{align}
where the black hole started in EPR pairs and time-evolved by $U$. Here $R$ is the early radiation, $C$ is the remaining black hole and $D$ is the outgoing mode. The goal is to find the partner $\widetilde{O}_{CR}^{T}$ of an arbitrary operator $O_{D}$ on the outgoing mode~\footnote{Construction of $\widetilde{O}_{CR}^{T}$ is not unique.}:
\begin{align}
(O_{D}\otimes I_{CR})|\Psi\rangle = (I_{D}\otimes \widetilde{O}_{CR}^{T}) |\Psi\rangle
\end{align}
or graphically
\begin{align}
\figbox{1.0}{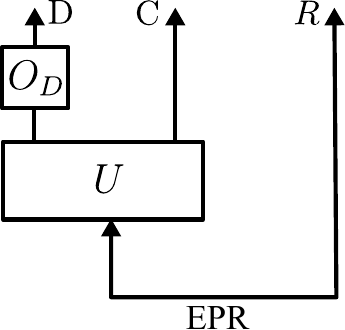} \ = \ \figbox{1.0}{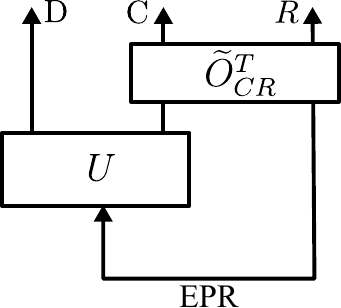} \label{eq:unsplit}
\end{align}
Existence of such partner operators is guaranteed since the mutual information $I(D,CR)$ is maximal. However, since $I(D,C)=0$, partners cannot be supported on $C$. That is, partner operators $\widetilde{O}_{CR}^{T}$ must have non-trivial support on $R$. We have seen that, from the perspective of the outside quantum mechanical description, this fact (the use of the early radiation $R$ in writing $\widetilde{O}_{CR}^{T}$) is the origin of the firewall problem (or relatedly the state-dependence problem and the non-locality problem).

Here our aim is to construct the partner $\widetilde{O}_{CR}^{T}$ by using as few qubits from $R$ as possible. To be explicit, let $A$ be a small subsystem of $R$, and $B$ be the complementary subsystem inside $R$ (so $R=AB$). Here we would like to construct a partner $\widetilde{O}_{CA}^{T}$ by using only qubits on $A$ from $R$, as graphically shown below:
\begin{align}
\figbox{1.0}{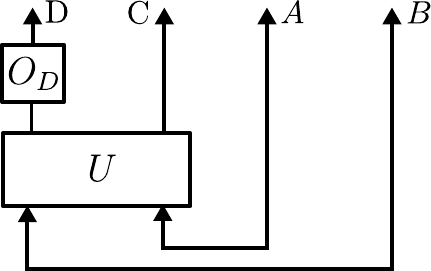} \ \approx \ \figbox{1.0}{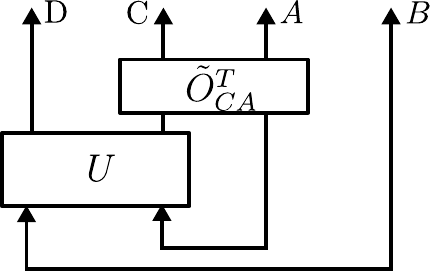} \label{eq:split}
\end{align}
where $R$ is split into two subsystems $AB$.

A key observation in~\cite{Beni18} is that the task of reconstructing $\widetilde{O}_{CR}^{T}$ for a given $O_{D}$ can be performed by using the Hayden-Preskill recovery protocol. Namely, by rotating the above figures in Eq.~\eqref{eq:split} by 180 degrees and bent some arrows, we obtain the following relation:
\begin{align}
\figbox{1.0}{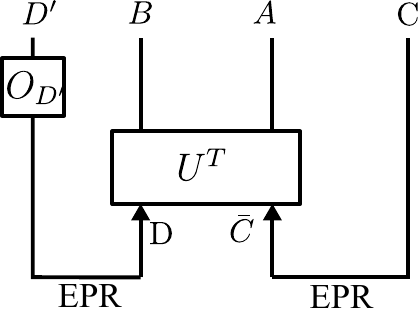} \ \approx  \figbox{1.0}{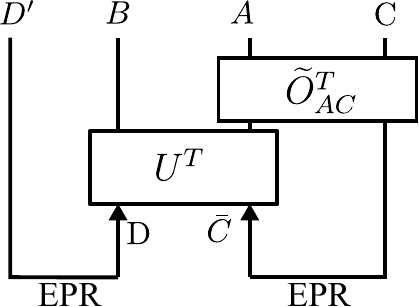} 
\end{align}
Here $U^{T}$ represents the transpose of $U$ as the diagram of $U$ is rotated upside down. In the rotated diagram, $D$ is interpreted as an ``input'' while its partner is constructed on the remaining black hole $C$ and just a little bit of qubits on $A$ from the early radiation $R$. It is immediate to see that the structure of the diagram is identical to the one for the Hayden-Preskill thought experiment. 

As such, the partner operator $\widetilde{O}_{CA}^{T}$ can be constructed by using the Hayden-Preskill recovery protocol from~\cite{Yoshida:2017aa}, as long as $A$ is bigger than $D$ and OTOCs between $A$ and $D$ decay~\footnote{The partners of interior operators are often called mirror operators in the literature. It is an interesting coincidence that Hayden and Preskill referred to their findings as ``black hole as mirror''.}.
It is worth summarizing the discussion so far as follows: 
\begin{align}
\text{HP recovery:}\qquad \text{input $A$} \qquad \text{reconstruction on $BD$}\\
\text{Interior operator:}\qquad \text{input $D$} \qquad \text{reconstruction on $CA$}.
\end{align}
We will study actual expressions of reconstructed interior operators more concretely in a future work.

\subsection{Young black hole}

The Hayden-Preskill recovery, as in the original Hayden-Preskill thought experiment, and the construction of interior operators outlined above can be achieved by using the same quantum information theoretic technique, namely the reconstruction protocol from~\cite{Yoshida:2017aa}. It is, however, worth emphasizing that these two phenomena, the Hayden-Preskill recovery and the construction of interior operators, are \emph{physically distinct}. In the original Hayden-Preskill thought experiment, the presence of the early radiation $R$, which is maximally entangled with the black hole, was essential for the recoverability of input quantum states. As such, the recovery was possible only for old black holes after the Page time. On the other hand, the aforementioned construction of the interior operators is possible for young black holes too as we will see below. In fact, the age of the black hole is not so essential. 

To see this explicitly, let us consider a young black hole which is not yet maximally entangled. One may model such a black hole simply as follows:
\begin{align}
\figbox{1.0}{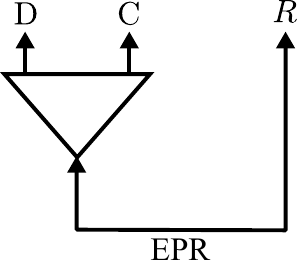} \ =\ \frac{1}{\sqrt{d_{R}}}\sum_{j=1}^{d_{R}} |j\rangle_{DC} \otimes |j\rangle_{R}
\end{align}
where the black hole Hilbert space $CD$ is $2^{n}$-dimensional (in a sense that the Bekenstein-Hawking entropy is $S_{\text{BH}}=n$) while the early radiation $R$ is supported on a $d_{R}$-dimensional Hilbert space with $d_{R}< 2^n$. The triangle in the figure denotes an isometric embedding, emphasizing that $DC$ is bigger than $R$. Taking $d_{R}=1$ would correspond to a one-sided black hole. The quantum state of interest is obtained by time-evolving the above state:
\begin{align}
|\Psi\rangle\ =\ \figbox{1.0}{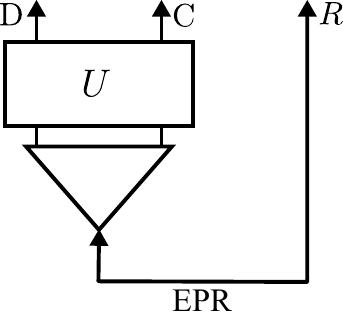}
\end{align}
For such a time-evolved young black hole, let us rotate the diagram to obtain
\begin{align}
\figbox{1.0}{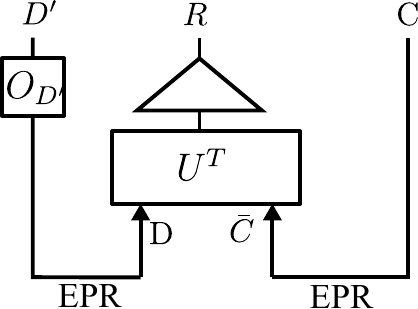}.
\end{align}
Then it is immediate to see that the partner can be reconstructed on $C$ and some small subsystem $A$ of $R$, as long as $A$ is bigger than $D$. Hence, even when a black hole is young, the interior operator can be reconstructed by using the Hayden-Preskill recovery protocol (as long as $R$ itself is bigger than $D$). 

Another (perhaps more illuminating) way to see that the reconstruction method works for a young black hole is to start with an old black hole, as in Eq.~\eqref{eq:old-state}, and project a subsystem of $R$ onto some fixed quantum state. Starting from a maximally entangled old black hole, let us split $R$ into $AB$ as before. Recall that we were able to reconstruct a partner operator $\widetilde{O}^{T}_{CA}$ on $CA$ without using $B$. Let us apply a projection onto some quantum state $|\psi^*\rangle_{B}$ supported on $B$~\footnote{Note $(I \otimes |\psi^*\rangle \langle \psi|)| \text{EPR}\rangle = |\psi\rangle \otimes |\psi^*\rangle$.}:
\begin{align}
\figbox{1.0}{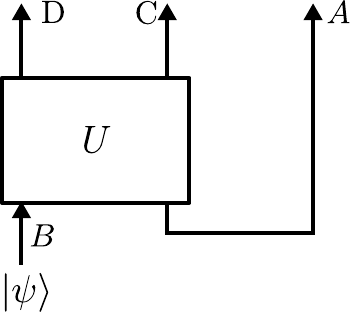}
\end{align}
After the projection, the system reduces to a young black hole which is entangled only with a small subsystem $A$. The reconstructed operator on $CA$ works as good as before the projection since its expression does not involve $B$ at all~\footnote{More precisely, it is because $D$ is (nearly) maximally entangled with $CA$ when OTOC between $A$ and $D$ decay. In quantum information theoretic term, the subsystem $B$ is said to be decoupled from $D$.}

From these observations, we notice that the aforementioned reconstruction of interior operators is \emph{fault-tolerant} against perturbations on $R$. Fault-tolerance is a quantum information theoretic term to say that encoding of quantum information is resilient against some types of errors. Here, the black hole interior operators are immune to rather drastic noises which would damage \emph{all the qubits} on $B\subset R$~\footnote{More precisely, it is an erasure error}. In qubits count, the reconstruction is robust even if any subsystem of $S_{\text{BH}} - \epsilon $ qubits from the early radiation are damaged as long as $\epsilon \gtrapprox |D|$. It is worth emphasizing that one can choose $A$ to be any subsystem of $R$ as long as OTOCs are small. Moreover, choosing different subsystems $A$ leads to different expressions of the interior operators $\widetilde{O}^{T}_{CA}$. The existence of multiple equivalent expressions of operators is a key characteristic of quantum error-correcting code. The necessity of such fault-tolerance was previously pointed out by Maldacena and Susskind~\cite{Maldacena13}. The above observation provides a physical account on how fault-tolerance is achieved due to scrambling dynamics of black holes.

So far we have shown that the partner operators can be constructed on $CA$ where $A$ is a small subsystem in $R$. In a sense, our construction is only ``slightly'' non-local as it uses only a small portion from the early radiation $R$. However, our construction does not resolve some of important problems identified in the context of the firewall problem. First, the reconstruction still requires an access to distant radiation $R$ albeit very little. Second, the reconstruction is state-dependent. For a generic initial state $(I\otimes K)|\Phi^{\text{EPR}}\rangle$, the recovery protocols should be run with $U \rightarrow UK^{*}$, so the reconstructed operators depend on $K$. Relatedly, the quantum circuit complexity of reconstruction can be huge depending on the complexity of $K$. We will remove these undesirable features in the next section by considering the effect of including the infalling observer.

\subsection{Monogamy and scrambling}

\begin{figure}
\centering
\includegraphics[width=0.27\textwidth]{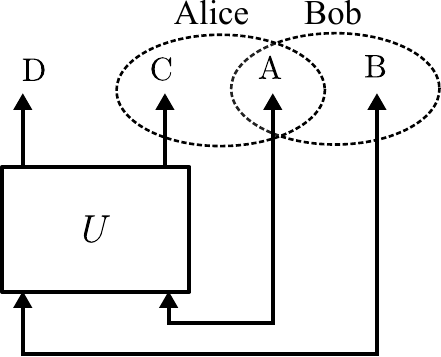} 
\caption{The AMPS puzzle from the perspective of the monogamy of entanglement. Alice can distill an EPR pair by accessing $AC$ while Bob can distill an EPR pair by accessing $AB$. Note that $A$ can be any subsystem of the early radiation $R$ as long as $|A|\gtrapprox |D|$ due to scrambling property of $U$.
}
\label{fig-fight}
\end{figure}

Our finding in this section sheds a new light on the AMPS puzzle through the lens of quantum information scrambling. The important lesson from the aforementioned reconstruction is that whoever possesses a tiny portion $A$ of the early radiation $R$ will be able to see some degrees of freedom which are entangled with the outgoing mode $D$. Namely, if $A$ is included to $C$ as degrees of freedom which Alice can access, then she is able to see an EPR pair between $D$ and $AC$. On the other hand, if $A$ is left untouched, the outgoing mode $D$ is entangled with $R=AB$ and Bob is able to see an EPR pair from the outside. See Fig.~\ref{fig-fight} for schematic illustration of the situation.

These statements can be made quantitative by using the monogamy relation of the mutual information. Decay of OTOCs implies that $I^{(2)}(D,AC)$ is nearly maximal. Since $I^{(2)}(D,AC)+I^{(2)}(D,B)=2 S_{D}^{(2)}$, this suggests that $I^{(2)}(D,B)$ is close to zero. Namely, Alice reconstructs $\overline{D}$ on $AC$ while Bob cannot reconstruct it on $B$.

It is worth emphasizing that $A$ can be any subsystem of $R$ as long as $|A|\gtrapprox |D|$~\footnote{A common misunderstanding is to think $A$ as qubits which are entangled with $D$ (i.e., the distilled qubits in the AMPS thought experiment). They are not. }. Hence, from the perspective of the infalling observer, it is rather ``easy'' to steal the EPR pair from the outside observer as she needs to take a few qubits in $R$ from Bob. We will make this intuition more precise in Section~\ref{sec:complexity} and Section~\ref{sec:outside}. In fact, we will show that generic perturbations to the black hole by the infalling observer will disentangle the outgoing mode $D$ from the early radiation $R$ \emph{without} ever accessing $R$. 

\section{State-independent interior operators}\label{sec:complexity}

In the previous section, we have pointed out that the interior operators can be constructed on the remaining black hole $C$, together with a little bit of qubits $A$ from the early radiation $R$. This construction, however, still suffers from the state-dependence and non-locality problems. 

In this section and the next, we discuss the effect of explicitly including the infalling observer to the firewall problem. We will show that inclusion of the infalling observer to the system, no matter how it is done, leads to significant gravitational backreaction which disentangles the outgoing mode $D$ from the early radiation $R$. This decoupling phenomena enables us to construct interior operators in a way independent of the initial state of the black hole without using any degrees of freedom from the early radiation $R$. Hence, the inclusion of the infalling observer enables state-independent interior operators and resolves the non-locality problem. Yet, the construction \emph{depends} on how the infalling observer is introduced to the system. In this sense, our construction is \emph{state-independent}, but \emph{observer-dependent}. 

In this section, we will demonstrate that a certain small perturbation to the black hole is sufficient to disentangle the outgoing mode from the early radiation. The main goal of this section is to simply point out that it is possible to disentangle the outgoing mode $D$ from the early radiation $R$ and construct the interior partner without using $R$. Discussions on physical interpretations of such a perturbation are given in the next section. Readers may find the particular scenario we consider in this section rather fine-tuned and artificial. In the next section, we will see that a similar phenomena occurs without any fine-tuning, in a rather universal manner, regardless of how perturbations are added to the black hole. 

\begin{figure}
\centering
\includegraphics[width=0.25\textwidth]{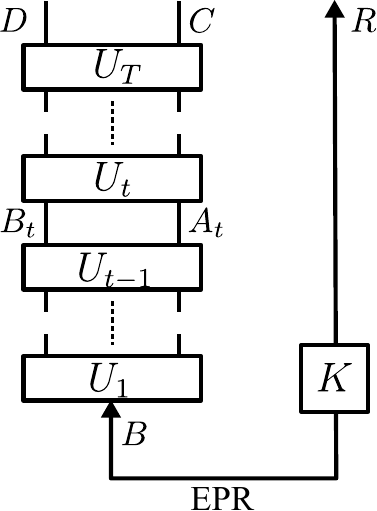} 
\caption{Discretized time evolution of a large AdS black hole with long throat. $K$ represents some complex unitary and $A_{t}$ are the confined modes. $R$ is far distant on the right-hand side. 
}
\label{fig-ADS-dynamics}
\end{figure}

\subsection{Long-throat AdS black hole}

Let us consider a large two-sided AdS black hole with long throat (i.e., $K$ is highly complex). The AdS assumption is used to ensure that the black hole does not evaporate. Let us imagine that the black hole initially starts with $(I\otimes K)|\Phi^{\text{EPR}}\rangle_{BR}$ and time-evolves by $U$ acting on $B$. Imagine that $U$ is discretized into small time steps:
\begin{align}
U = U_{T}\cdots U_{1}.
\end{align}
We emphasize that details of the discretization are not important in the following discussion. The quantum state at $t=T$ is depicted in Fig.~\ref{fig-ADS-dynamics}. Here intermediate Hilbert spaces are labelled by $A_{t}, B_{t}$ where $A_{t}$ corresponds to the modes on the AdS boundary while $B_{t}$ corresponds to other degrees of freedom including modes between the boundary and the stretched horizon as well as other entities living at the stretched horizon. 

\begin{figure}
\centering
(a)\includegraphics[width=0.25\textwidth]{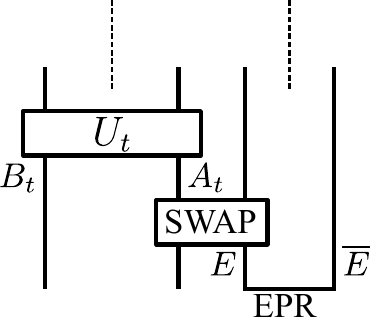} \qquad
(b)\includegraphics[width=0.25\textwidth]{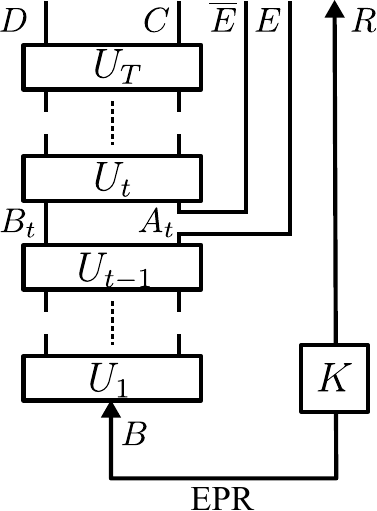} 
\caption{A protocol for an infalling observer to access $A_{t}$. EPR pair is prepared on $E\overline{E}$. (a) SWAP operation between $A_{t}$ and $E$. (b) The outcome. $\overline{E}$ and $A_{t}$ are entangled.  
}
\label{fig-ADS-SWAP}
\end{figure}

We now show that the partner of interior operators $D$ can be reconstructed by quantum operations which are strictly localized on the left-hand side without ever accessing $R$. Consider the quantum state of the black hole at time $t$:
\begin{align}
 ( U_{t-1}\cdots U_{1} \otimes K) |\text{EPR}\rangle_{BR} .
\end{align}
To begin, we append two subsystems $E,\overline{E}$, which have the same dimensionality as $A_{t}$ at time $t$, and prepare EPR pair on $E\overline{E}$ as shown in Fig.~\ref{fig-ADS-SWAP}(a). The resulting state can be written explicitly as follows:
\begin{align}
|\Psi(t)\rangle = ( U_{t-1}\cdots U_{1} \otimes K) |\text{EPR}\rangle_{BR}  \otimes |\text{EPR}\rangle_{E\overline{E}}.
\end{align}
Note that the EPR pair on $E,\overline{E}$ is added to the black hole Hilbert space by hand, and is not a part of the original Hilbert space. Let us apply SWAP operator between $A_{t}$ and $E$ as shown in Fig.~\ref{fig-ADS-SWAP}(a):
\begin{align}
|\Psi'(t)\rangle = \text{SWAP}_{A_{t}\overline{E}}\ \big(  ( U_{t-1}\cdots U_{1} \otimes K) |\text{EPR}\rangle_{BR}  \otimes |\text{EPR}\rangle_{E\overline{E}} \big).
\end{align}
Here $\text{SWAP}$ is a unitary operator which acts as $\text{SWAP}(|i\rangle\otimes |j\rangle) = |j\rangle \otimes |i \rangle$. 

The outcome of these operations is depicted in Fig.~\ref{fig-ADS-SWAP}(b). We find that $\overline{E}$ is entangled with $A_{t}$. Namely, we notice that $\overline{E}$ can play the role of $A$ in section~\ref{sec:reconstruction}. Hence, the interior operators (the partner of operators on $D$) can be reconstructed on $\overline{E}C$ by applying the recovery unitary from the Hayden-Preskill recovery protocol with $U'=U_{T}\cdots U_{t}$ as long as $T-t \gtrapprox t_{\text{scr}}$. More explicitly, at time $t=T$, the quantum state of the black hole is given by
\begin{align}
|\Psi'(T)\rangle = \big( U_{T}\cdots U_{t} \otimes I_{E\overline{E}R} \big)  \ \text{SWAP}_{A_{t}\overline{E}}\ \big( ( U_{t-1}\cdots U_{1} \otimes K) |\text{EPR}\rangle_{BR}  \otimes |\text{EPR}\rangle_{E\overline{E}} \big).
\end{align}
Given an arbitrary operator $O_{D}$ on $D$, one can construct a partner operator $\widetilde{O}^{T}_{C\overline{E}}$ on $C\overline{E}$ such that 
\begin{align}
(O_{D} \otimes I )|\Psi'(T)\rangle \approx (I \otimes \widetilde{O}^{T}_{C\overline{E}} )|\Psi'(T)\rangle.
\end{align}
Due to the additional of the $E\overline{E}$ system and the SWAP operation, the outgoing mode $D$ is no longer entangled with the early radiation $R$. Namely, the partner $\widetilde{O}^{T}_{C\overline{E}}$ can be reconstructed without involving the early radiation $R$ ever. 

\subsection{State-independence}

Having constructed a partner operator without using the early radiation $R$, let us discuss the problem of the state-dependence. As is evident from Fig.~\ref{fig-ADS-SWAP}(b), the reconstruction of interior operators depends only on $U'= U_{T}\cdots U_{t}$. Importantly, it has no dependence on the unitary $K$, and hence is state-independent. As such, the reconstruction is not only efficient with $O(|T-t|)$ quantum circuit complexity, but also independent from the choice of the initial state $(I\otimes K)|\Phi^{\text{EPR}}\rangle_{BR}$. 

The aforementioned protocol works for the case of one-sided AdS black holes. Let us represent the initial state of the black hole by $K |0\rangle^{\otimes n}$ and time-evolve it by $U=U_{T}\cdots U_{1}$. This is equivalent to projecting the right-hand side $R$ in Fig.~\ref{fig-ADS-dynamics} onto $|0\rangle^{\otimes n}$. Since the recovery protocol does not depend on any quantum operation on $R$, be it unitary or non-unitary, the reconstructed interior operators are independent from the initial state
~\footnote{When the initial state of the black hole is given by $n$ copies of EPR pairs, projecting $R$ onto $|\psi^*\rangle$ will project the black hole on $B$ onto $|\psi\rangle$. Naively, this appears to suggest that it might be possible to influence the black hole on $B$ and its interior by non-unitary operation on $R$, contradicting with our claim. The important point is that the $2^n$-dimensional subspace where the black hole is initially entangled with $R$ should be interpreted as a coarse-grained Hilbert space with roughly equal energy. As long as projection on $R$ picks some quantum states of the black hole in thermal equilibrium within the typical energy window, reconstructed interior operators will behave well. Yet, if projection on $R$ brings the black hole on $B$ to a quantum state with much higher/lower temperatures, then reconstruction won't work well. See section~\ref{sec:code} for further discussion on this point. 
}. 
The important point for us is that OTOCs between $A$ and $D$ decay for one-sided black holes too.

\begin{figure}
\centering
\includegraphics[width=0.4\textwidth]{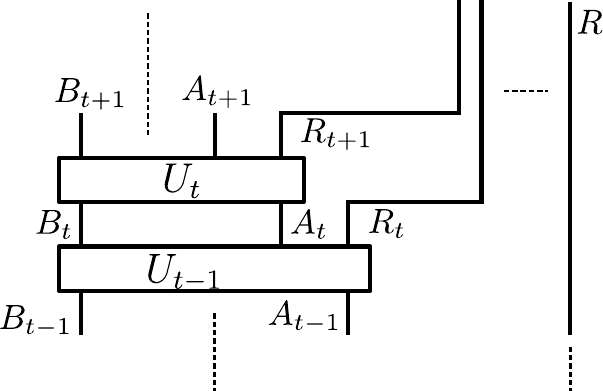} 
\caption{The dynamics of an evaporating black hole. $R_{t}$ represents the modes which escape and join $R$ while $A_{t}$ are the confined modes.
}
\label{fig-evaporation}
\end{figure}

Finally let us return to the discussion of evaporating black holes as originally considered by AMPS. We model the evaporation dynamics as a sequential application of the procedures depicted in Fig.~\ref{fig-evaporation}. Here $A_{t}$ are the confined mode inside the zone whereas $R_{t}$ are the modes which escape from the zone and never return. As time passes, $R_{t}$ joins the radiation $R$, and the entropy of the black hole decreases~\footnote{The AdS black hole can be viewed as a limit where $R_{t}$ disappears.}. The aforementioned protocol works successfully by swapping $E$ and $A_{t}$. Here it is crucial for an infalling observer $E$ to interact with the confined modes $A_{t}$ rather than the escaping modes $R_{t}$ as $R_{t}$ is completely decoupled from $A_{t}B_{t}$. Again, the interior operators can be reconstructed in a state-independent and low-complexity manner without ever accessing $R$.  

Readers might find that the above construction of the interior operator is rather artificial and fine-tuned. Obviously, this setup was devised in order to mimic the situation considered in Section~\ref{sec:reconstruction}. One may interpret $E\overline{E}$ as certain measurement apparatus or physical probe which is introduced to the black hole Hilbert space~\footnote{Situations analogous to the $E\bar{E}$-system plus SWAP can be realized by performing informationally complete measurements (POVMs). }. In the next section, we will consider corresponding physical situations and how our reconstruction resolves the AMPS puzzle. We will also argue that the underlying physical phenomenon is quite universal in gravitational systems and does not require fine-tuning. 

\subsection{Code subspace}\label{sec:code}

We have presented state-independent reconstruction of interior operators that are completely insensitive to perturbations on the early radiation $R$. Strictly speaking, this conclusion on state-independence is a bit of exaggeration. For instance, large perturbations added to $A_{t}$ will influence the Hamiltonian under which $U_{t}$ evolves. As such, it is important to note that the state-independence is an approximate statement. 

To understand the validity of the approximation, let us return to the underlying assumption behind using simplified descriptions of quantum black holes as $S_{\text{BH}}$-qubit systems. In static geometries, the Bekenstein-Hawking entropy $S_{\text{BH}}$ can be interpreted as the coarse-grained entropy. Namely it suggests that there are $\approx 2^{S_{\text{BH}}}$-dimensional Hilbert space, denoted by $\mathcal{H}_{\text{code}}$, which have an identical classical black hole geometry. This subspace is spanned by quantum states which correspond to black holes in thermal equilibrium with the same thermodynamic parameters, such as mass $M$, angular momentum $J$ and charge $Q$. Those black hole microstates, however, differ in a subtle manner. They may differ in terms of matter content on a fixed geometry and other high energy objects living at the stretched horizon. Black holes in thermal equilibrium with the same background classical geometry span the subspace $\mathcal{H}_{\text{code}}$.

The very motivation in reconstructing interior operators in a state-independent manner is to understand quantum field theory on a curved space-time experienced by the infalling observer. This suggests that reconstructed operators must be state-independent on a fixed black hole geometry, but can be dependent on choices of geometries. Our claim on the state-independence is restricted to the validity of $S_{\text{BH}}$-qubit toy models, and hence is an approximate statement inside the subspace $\mathcal{H}_{\text{code}}$ which is determined by the classical black hole geometry.

These interpretations become sharp in the context of the AdS/CFT correspondence. In~\cite{Almheiri:2015ac}, Almheiri, Dong and Harlow proposed that the subspace $\mathcal{H}_{\text{code}}$ can be interpreted as a codeword subspace of a quantum error-correcting code. Their insight, as well as concrete toy model of such a scenario~\cite{Pastawski15b}, suggests that bulk operators can be interpreted as logical operators which label different codeword states in $\mathcal{H}_{\text{code}}$~\footnote{
When perturbations being the system completely outside $\mathcal{H}_{\text{code}}$, the Hayden-Preskill recovery phenomenon may not be possible as discussed in~\cite{Yoshida:2019aa}.
}. 

From the perspective of the boundary Hilbert space, one may also interpret the origin of the codeword subspace $\mathcal{H}_{\text{code}}$ by relying on the Eigenstate Thermalization Hypothesis (ETH)~\cite{Srednicki:1994aa}. In particular, for strongly interacting quantum many-body systems, eigenstates from a tiny energy window are expected to look thermal in small subsystems. One may think that the subspace $\mathcal{H}_{\text{code}}$ is spanned by these wavefunctions contained in the small energy window. 

Keeping this caveat in mind, our claim in a more precise term is that the reconstruction of interior operators is fault-tolerant against perturbations added to the entangled partner $R$ as well as any quantum operations added before time $t$, provided that the black hole is initially in thermal equilibrium~\footnote{By thermal equilibrium, we mean that time-ordered correlation functions reach their stationary values.}. We also believe that the above claim includes cases of evaporating black holes where the system reaches thermal equilibrium quickly. The typical time scale for our reconstruction procedure is the scrambling time which is of order $\approx r_{S}\log r_{S}$ in the Schwartzshild black hole whereas the thermalization time for small perturbation is of order $\approx r_{S} \approx \frac{1}{T}$. 

\section{Backreaction by infalling observer}\label{sec:outside}

In this section, we show that gravitational backreaction by the infalling observer, who simply fall into a black hole, disentangles the outgoing mode $D$ from the early radiation $R$. As a result, the infalling observer crosses the horizon smoothly and observes quantum entanglement between the outgoing mode $D$ and the interior mode $\overline{D}$. We also show that the interior mode $\overline{D}$ is insensitive to any perturbations to the early radiation $R$ from geometric perspective. Our conclusions will be supported quantitatively by using decay of OTOCs. 

\subsection{Including Alice}\label{sec:outside_Alice}

\begin{figure}
\centering
(a) \includegraphics[width=0.35\textwidth]{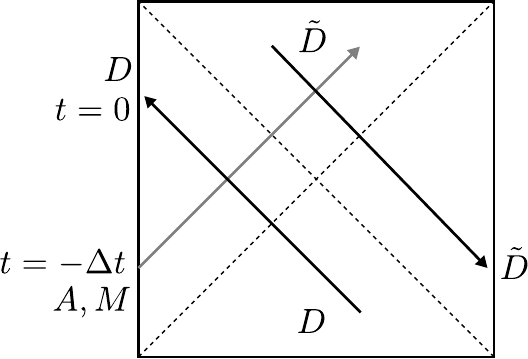} 
(b) \includegraphics[width=0.322\textwidth]{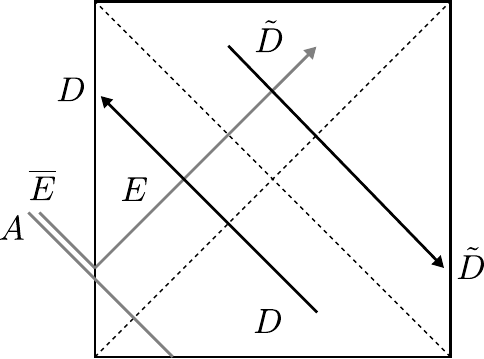} 
\caption{(a) The effect of an apparatus $M$. (b) The protocol from section~\ref{sec:complexity}.
}
\label{fig-apparatus}
\end{figure}

Our reconstruction protocol required an access to the earlier mode. Readers might think that we have just traded the spatial non-locality with the temporal non-locality. The infalling observer can access earlier modes by simply falling into the black hole earlier. To be concrete, let us focus on the two-sided eternal AdS black hole. Let $D$ be the outgoing mode that reaches the boundary at $t=0$ as shown in Fig.~\ref{fig-apparatus}(a). As an infalling observer, one may consider some apparatus $M$ which departs the boundary at $t = -\Delta t$. To minimize the effect of adding $M$, one may consider a limit where $M$ assumes the smallest possible energy, say $\approx 1/\beta$. Here the apparatus $M$ travels along the infalling mode $A$ at $t= -\Delta t $. 

To understand the effect of $M$, it is convenient to draw the partner mode $\tilde{D}$ in the Penrose diagram. The reason why we used $\tilde{D}$ instead of $\overline{D}$ will become clear later. Here the trajectory of $\tilde{D}$ can be constructed by rotating the trajectory of $D$ by 180 degrees as in Fig.~\ref{fig-apparatus}(a). Near the horizon, $D$ and $\tilde{D}$ can be viewed as Rindler modes in the opposite regions. Near the boundary, $D$ and $\tilde{D}$ can be viewed as entangled degrees of freedom in the boundary conformal field theory which is in the thermofield double state. In the absence of the apparatus $M$, Bob can verify the entanglement between $D$ and $\tilde{D}$ by simply measuring the correlation functions between them. Also, one may explicitly compute the mutual information $I(D,\tilde{D})$ by using the Ryu-Takayanagi formula~\cite{Ryu06}. Here the interior operator $\tilde{D}$ is constructed exclusively on the early radiation $R$ (or degrees of freedom in the right hand side).  

However, in the presence of the measurement apparatus $M$, Bob's quantum entanglement between $D$ and $\tilde{D}$ is disturbed. Precise form of interactions between the apparatus $M$ and other degrees of freedom is not so important. In fact, the interaction is universal in a sense that it is a gravitational effect. Namely, since $M$ becomes a gravitational shockwave which shifts the geometry near the horizon, the correlation function between $D$ and $\tilde{D}$ decays. The two-point correlation functions between $D$ and $\tilde{D}$ in the presence of a shockwave is identical to OTOCs defined for a thermal state~\cite{shenker2014black}. As such, when $\Delta t \gtrapprox t_{\text{scr}}$, Bob can no longer verify the entanglement between the outgoing mode $D$ and the early radiation $R$ by measuring $D$ and $\tilde{D}$. This hints that Bob's quantum entanglement is destroyed by the infalling observer who has crossed the horizon. 

One might think that this observation does not fully deny a possibility that the outgoing mode $D$ is entangled with some other degrees of freedom in the early radiation $R$. In order to make a more quantitative argument, let us explicitly study the effect of adding the apparatus $M$. Suppose that the quantum state of a black hole is given by EPR pairs $|\Phi^{\text{EPR}}\rangle_{ABR}$ at $t = - \Delta t$. We then append the subsystem for the apparatus $M$ to the black hole. We assume that $M$ starts with some particular pure state, say $|0\rangle$. Under the time-evolution by a unitary operator $U$, we obtain the following state:
\begin{align}
(U_{BAM}\otimes I_{R}) (|0\rangle_{M} \otimes |\Phi^{\text{EPR}}\rangle_{BAR}) = \figbox{1.0}{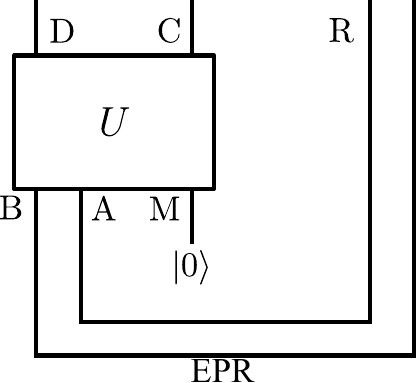}\ . \label{eq:apparatus}
\end{align}
Here we increased the Hilbert space dimension of the black hole in order to account for the increase of the coarse-grained entropy due to the inclusion of $M$~\footnote{The increase of the entropy is approximately given by $E/T$ where $E$ is the energy of $M$ including its rest mass and $T$ is the temperature.}. We assume that this extension of the Hilbert space via equilibration occurs in thermalization time which is much shorter than the time scale of applying $U$ (or scrambling time). If $U$ is Haar random unitary operator, one can show that $I(C,D)$ becomes nearly maximal when $|M| \gtrapprox |D|^2 $. This suggests that quantum entanglement between $D$ and $R$, which Bob would have seen, disappears as a result of including $M$. We emphasize that this argument on $I(C,D)$ is not limited to the AdS/CFT correspondence as it only requires the quantum state of a black hole to be maximally entangled. 

A similar setting was previously considered by Verlinde and Verlinde in a slightly different context~\cite{Verlinde:2013aa}, and we will make further comments on their work in appendix~\ref{sec:Verlinde}. 

While the aforementioned estimate assumes that $U$ is a Haar random unitary, one can relate certain average of OTOCs to the mutual information $I(C,D)$ for the state defined in Eq.~\eqref{eq:apparatus}. Again, generic decay of OTOCs is sufficient to show that $I(C,D)$ is nearly maximal~\footnote{This statement can be shown by a slight generalization of results from~\cite{Hosur:2015ylk, Yoshida:2017aa}. }. Also, further analysis with energy conservation into consideration suggests that $|M| \gtrapprox |D|$ is sufficient to restore nearly maximal correlation in $I(C,D)$ as discussed in appendix~\ref{sec:interaction}, 

\subsection{Sending probes}

In the previous subsection, we discussed the effect of including the infalling observer (or a measurement apparatus). 
The situation, however, differs from the reconstruction method in section~\ref{sec:complexity} in a subtle detail. Also we had to change the size of the black hole Hilbert space by including $M$. Here we consider a situation which exactly mimics our reconstruction method.  

Instead of adding the apparatus $M$, we think of replacing the infalling mode $A$ at $t=-\Delta t$ with some ``probe'' mode. Let us prepare the EPR pair on $E\overline{E}$. Let the outgoing mode $A$ at $t=-\Delta t$ escape from the boundary and replace the infalling mode $A$ with $E$ while keeping $\overline{E}$ as a reference qubit. See Fig.~\ref{fig-apparatus}(b) for the illustration. Here the earlier outgoing mode $A$ and the reference $\overline{E}$ remain outside the boundary whereas $E$ probes the bulk. This is essentially identical to the protocol with SWAP operation discussed in section~\ref{sec:complexity}. Decay of OTOCs implies that $I^{(2)}(D,\overline{E}C)$ becomes nearly maximal which suggests that the outgoing mode $D$ and $\overline{E}C$ are indeed entangled. This also implies that $D$ is no longer entangled with $R$ due to the monogamy of quantum entanglement. Instead, it is entangled with the interior mode $\overline{D}$ which we constructed on Section~\ref{sec:complexity}. Again, this argument is not particularly limited to the AdS/CFT correspondence.

We have seen that including $E\overline{E}$ also disentangled the outgoing mode $D$ from the early radiation $R$. One merit of introducing the probe mode $E\overline{E}$ as in section~\ref{sec:complexity} is that the size of the Hilbert space for the black hole itself does not change. Note that this protocol differs from the scenario in the previous subsection where an apparatus $M$ was introduced. Nevertheless, we believe that these differences, which result from how we include the apparatus $M$ in the black hole Hilbert space, do not affect our main conclusion significantly. Some readers may find inclusion of extra objects, such as the apparatus $M$ and the ancilla qubits $\overline{E}E$, uncomfortable. The reason why we need to introduce extra degrees of freedom is that a black hole is treated as a system of discrete finite-dimensional qubits, and hence, there is no room for adding probes. In a more realistic field theoretic formulation, one may simply turn on the source term which creates the ``apparatus'' or the ``probe mode'' as an excitation. 

Regardless of how the probes or the apparatus are introduced, what is important for us is that OTOCs between $M$ (or $A$) and $D$ decay after the scrambling time. From gravitational perspective, this is a universal phenomenon resulting from the surface gravity and shockwave geometries. From quantum information theory perspective, this is sufficient to prove that the outgoing mode $D$ is disentangled from the early radiation $R$ and $\overline{D}$ can be reconstructed on $A$ and $C$ without involving $R$. Hence, the disentangling phenomenon occurs universally no matter how she falls~\footnote{One might wonder whether an observer can be introduced in a way which introduces no backreaction. From the outside quantum mechanical viewpoint, this appears to be not possible if the perturbations are added on one side of the black hole. Our argument relies on decay of OTOCs which is a rather generic feature of interacting quantum many-body systems. It is in principle possible that the value of OTOCs returns to $O(1)$ value after a very long time (of the order the quantum recurrence time). But in time scales which are relevant to the AMPS puzzle, we believe that OTOCs decay monotonically to a small stationary value and then oscillates around it. We believe that such a conclusion can be mathematically proven by utilizing ETH. One possible way to undo the effect of backreaction is to perform the Hayden-Preskill recovery protocol by accessing both sides of the black hole as we discuss later.}. 

Finally we speculate on non-uniqueness of reconstruction of interior operators. From the perspective of the outside observer, the interior operator is realized as $\tilde{D}$ supported on $R$. On the other hand, from the perspective of the infalling observer who travels along the infalling mode $A$, the interior operator is realized as $\overline{D}$ which is supported on the remaining black hole $C$ and the infalling mode $A$. Hence, we propose that reconstruction of $\overline{D}$ from section~\ref{sec:complexity} corresponds to quantum mechanical operators for the infalling observer who departed at $t= -\Delta t$~\footnote{
Although we have included an observer or an apparatus, one can discuss the monogamy of entanglement purely from the operator algebraic perspective. The important point is that representations of basis operators for $\overline{D}$ on $AC$ and $AB$ do not necessarily commute. 
}. In principle, the infalling observer may choose to fall into a black hole at any time by following any trajectory as long as it does not violate the causality. We speculate that different infalling trajectories correspond to different reconstructions of interior operators. An explicit relation between trajectories and reconstructions, however, is beyond the scope of this paper~\footnote{In the aforementioned scenarios, infalling observers were introduced to the system by appending ancilla Hilbert spaces. Readers might wonder if addition of ancilla systems are necessary to achieve the disentangling phenomena or not. One can actually avoid the use of ancilla systems simply by considering a more physical realistic description of black holes, at least within the context of the AdS/CFT correspondence. Let $\mathcal{H}_{\text{CFT}}$ be the full Hilbert space of the CFT and consider a black hole at temperature $\beta$ which lives on a subspace $\mathcal{H}_{\beta} \subset \mathcal{H}_{\text{CFT}}$ consisting of microstates at the same temperature. Inclusion an infalling observer or measurement apparatus increases the entropy of the black hole by $\Delta S = \frac{E}{T}$. The point is that physics of adding infalling observers can be discussed in the full CFT Hilbert space $\mathcal{H}_{\text{CFT}}$.}.

\subsection{Geometric interpretation}

We have argued that the outgoing mode $D$ is disentangled from the early radiation $R$ (or the right hand side) due to the decay of OTOCs caused by the infalling observer. Our primary focus was on assessing the effect of backreaction from the outside quantum mechanical description. Here we present geometric interpretation of our discussion~\footnote{I thank Xiaoliang Qi and Ying Zhao for useful discussions on this.}. 

\begin{figure}
\centering
\includegraphics[width=0.27\textwidth]{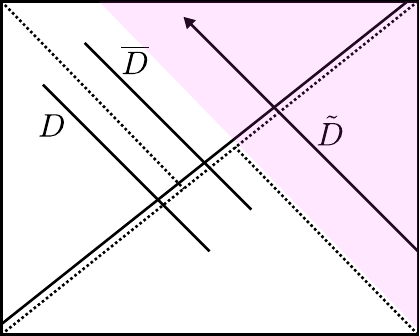} 
\caption{The infalling observer as a gravitational shockwave. The interior mode $\overline{D}$ is outside a shaded region that may be influenced by quantum operations on the right hand side. The original mode $\tilde{D}$ emerges from the right hand side, and is not longer entangled with $D$.
}
\label{fig-causal}
\end{figure}

For simplicity of discussion, let us consider the AdS eternal black hole. Given the outgoing mode $D$, one possible representation of interior partner operators can be constructed by time-evolving a corresponding mode on the right hand side. This operator, constructed exclusively on the degrees of freedom on the right hand side, was denoted by $\tilde{D}$. We now include the effect of the infalling observer $M$ as a gravitational shockwave and draw the backreacted geometry where the horizon is shifted as depicted in Fig~\ref{fig-causal}. 
If the time separation between the outgoing mode $D$ and the infalling observer $M$ is longer than or equal to the scrambling time, the interior mode $\overline{D}$ can be found across the horizon and is outside the causal influence of any operations on the right hand side. This is the geometric explanation of the fault-tolerance and state-independence of the interior operator $\overline{D}$. It is worth noting that a similar observation applies to generic black holes in thermal equilibrium
~\footnote{Here we treated the effect of the perturbation by Alice first on the left hand side, and then discussed the effect of the perturbations added on the right hand side. One might wonder why we did not treat the effect of the perturbation on the right hand side first. According to the structure of quantum entanglement implied by decay of OTOCs, we should treat the effect of the perturbation on the left hand side first when addressing the experience of the infalling observer from the left hand side.}. Here it is important to note that the interior mode $\overline{D}$ in a backreacted geometry is different from the original partner operator $\tilde{D}$. 

This geometric interpretation enables us to deduce the experience of the infalling observer. From the backreacted geometry, we see that the infalling observer can cross the horizon smoothly and observe the entangled pair $D$ and $\overline{D}$. According to the causal structure, after observing the interior mode $\overline{D}$, Alice eventually enters a region which may be affected by quantum operations on the right hand side. Inside this region, shaded in Fig~\ref{fig-causal}, Alice may be hit by a gravitational shockwave created by the outside observer, according to the backreacted geometry. It is, however, unclear to us whether perturbations on the right hand side can actually influence the experience of the infalling observer at all in the absence of non-local interactions between the left and right hand sides (or between the remaining black hole and distant early radiation). In fact, we do not think that perturbation on $R$ influences Alice's infalling experience at all if it is strictly localized on $R$. 
In any cases, quantum operations on the right hand side do not affect Alice's experience of crossing a smooth horizon and seeing $D$ and $\overline{D}$. 

\section{On definition of state-independence}\label{sec:FAQ}

Some previous studies suggest that black hole interior operators must be state-dependent. In this section, we revisit these no-go arguments for state-independence and discuss how our construction avoids them. The main message is that our construction does not depend on the initial state of the black hole, but does depend on how the infalling observer is introduced, being \emph{state-independent}, but \emph{observer-dependent}.

\subsection{No-go arguments}

Let us briefly recall some versions of the no-go arguments for state-independence for an old black hole. Consider a maximally entangled black hole: 
\begin{align}
|\Psi\rangle = (I \otimes K) |\text{EPR}\rangle_{CDR}
\end{align}
where $C$ is the remaining black hole, $D$ is the outgoing mode and $R$ is the early radiation. Here $R$ can be decomposed into $\overline{C}$ and $\overline{D}$ such that $C$ and $D$ are entangled with $\overline{C}$ and $\overline{D}$ respectively. Given a unitary operator $O_{D}$ acting on $D$, a partner operator $\tilde{O}_{CR}$ must satisfy 
\begin{align}
(O_{D} \otimes I_{CR})|\Psi\rangle \approx (I_{D} \otimes \tilde{O}_{CR})|\Psi\rangle. \label{eq:partner1}
\end{align}
One possible representation can be constructed as follows:
\begin{align}
\tilde{O}_{CR} = K O_{\overline{D}}^{T} K^{\dagger}
\end{align}
by using the fact $(O_{D}\otimes I_{C\overline{CD}})|\text{EPR}\rangle = (O_{\overline{D}}^T\otimes I_{CD\overline{C}}) |\text{EPR}\rangle$.
However, this construction clearly depends on the initial state of the black hole. 

Next, let us recall the argument for a young black hole where the early radiation $R$ is absent. Consider a ``typical'' pure quantum state $|\Psi_{\text{young}}\rangle_{CD}$. For the sake of discussion, one may assume that $|\Psi_{\text{young}}\rangle_{CD}$ is drawn randomly from the Haar measure. If $|D|\ll |C|$, the mutual information $I(C,D)$ is nearly maximal (for typical states). Hence, given a unitary operator $O_{D}$, a partner operator $\tilde{O}_{C}$ does exist:
\begin{align}
( \tilde{O}_{C} \otimes O_{D} ) |\Psi_{\text{young}}\rangle_{CD} \approx |\Psi_{\text{young}}\rangle_{CD}. \label{eq:partner2}
\end{align}
For a given $|\Psi_{\text{young}}\rangle_{CD}$ and $O_{D}$, let us fix $\tilde{O}_{C}$ and ask if $\tilde{O}_{C}$ is also a partner operator for some other typical pure quantum state $|\Psi_{\text{young}'}\rangle_{CD}$. The answer is clearly no since $C$ and $D$ are entangled in a way which depends on each quantum state. 

A similar conclusion can be also obtained by considering an average over typical quantum states. Observe the following relation
\begin{align}
\int d |\Psi\rangle \   \langle \Psi |_{CD} ( \tilde{O}_{C} \otimes O_{D} ) |\Psi\rangle_{CD} = \frac{1}{2^n} \Tr(\tilde{O}_{C}\otimes O_{D})
\end{align}
where the integral is taken over the Haar measure. If $\tilde{O}_{C}$ were partner operators of $O_{D}$ for $|\Psi\rangle_{CD}$, we would have $\langle \Psi |_{CD} ( \tilde{O}_{C} \otimes O_{D} ) |\Psi\rangle_{CD}\approx 1$. But the RHS becomes zero for traceless unitary operators, such as Pauli operators. Hence one cannot choose the partner operators in a state-independent manner~\footnote{Arguments along this line are also used as an evidence for firewalls in typical black hole microstates~\cite{Marolf:2013aa}.}. 

\subsection{Observer-dependence}

Our construction of the black hole interior operators avoids the above no-go arguments simply due to the fact that it does not give partner operators for the initial state of the black hole, but for a slightly perturbed state after the inclusion of the infalling observer. To see this point explicitly, let us recall how the interior operators were reconstructed after including the infalling observer, specifically the scenario discussed in section~\ref{sec:outside_Alice}. Let $|\Phi(0)\rangle$ be the initial state of the black hole, which can be two-sided or one-sided (old or young). In section~\ref{sec:outside_Alice}, we explicitly appended an infalling observer to the system:
\begin{align}
|\Phi'(0)\rangle = |\Phi(0)\rangle \otimes |0\rangle_{M}. \label{eq:observer}
\end{align}
After the inclusion of the infalling observer, the whole system time-evolves by some chaotic unitary $U$:
\begin{align}
|\Phi'(t)\rangle = U(t) |\Phi'(0)\rangle. 
\end{align}
What we have constructed, by the use of the Hayden-Preskill recovery protocol, are the interior partner operators for the quantum state $|\Phi'(t)\rangle$, not for the unperturbed original quantum state $|\Phi(t)\rangle$. As such, previous no-go arguments do not apply to our construction. 

It is, however, worth emphasizing that we are not entirely free from dependence on specifics of the system. Namely, our construction manifestly depends on how the infalling observer is introduced to the system. For instance, if a different initial state of the infalling observer is chosen (say $|1\rangle$) in Eq.~\eqref{eq:observer}, the expression of interior operators would be different. Hence, while our construction of interior operators does not depend on the initial state of the black hole $|\Phi(0)\rangle$, it does depend on \emph{observers}.

Similar justifications hold for other scenarios of introducing infalling observers to the black hole. For instance, we have illustrated the use of an ancilla mode $\overline{E}$ as a physical probe to the black hole. In this case, the reconstructed interior operators are of course the ones for perturbed states with such ancilla modes. It is unclear to us how to treat the inclusion of the infalling observers on a unified footing. But formally, this procedure can be expressed as an isometric quantum operation:
\begin{align}
|\Phi'(0)\rangle = \Omega( |\Phi(0)\rangle ).
\end{align}
Here the isometry $\Omega$ accounts for all the quantum operations associated with the inclusion of the infalling observer. Generally speaking, $\Omega$ consists of some simple (low complexity) unitary operations as well as addition of ancilla systems. Obviously, $|\Phi'(0)\rangle$ depends on the choice of the initial state $|\Phi(0)\rangle$ as well as how the infalling observer is introduced, namely the map $\Omega$. After the inclusion of the infalling observer, the black hole time-evolves by some chaotic dynamics: $|\Phi'(t)\rangle = U(t) |\Phi'(0)\rangle$. Our claim is that construction of interior operators for such a slightly perturbed black hole $|\Phi'(t)\rangle$ does not depend on the initial state $|\Phi(0)\rangle$, but depends on the map $\Omega$. Discussions on actual expressions of interior operators for each possible scenario of introducing infalling observers will be given in a future work.

\section{Resolution of AMPS puzzle}\label{sec:puzzle}

In this section, we present a resolution of the AMPS puzzle.

\subsection{AMPS thought experiment}

Let us consider a version of the AMPS thought experiment. Given an old black hole that is maximally entangled with the early radiation $R$, consider an outgoing mode $D$ which has been just emitted from the black hole. The outgoing mode $D$ is entangled with some degrees of freedom in the early radiation $R$. The outside observer Bob distills a qubit (or qubits) $\tilde{D}$ that is entangled with the outgoing mode $D$ by performing some careful quantum operation on $R$. Bob may isolate the qubit $\tilde{D}$ from the black hole or hand it to Alice who is about to fall into a black hole. After crossing the horizon smoothly, Alice will see an interior mode $\overline{D}$ which is entangled with the outgoing mode $D$. This, however, leads to a contradiction because $D$ is also entangled with $\tilde{D}$. 

The fallacy of the above argument is clear. When Alice falls into a black hole, the outgoing mode $D$ is disentangled from the distilled qubit $\tilde{D}$ due to the backreaction which makes OTOCs decay. Furthermore, Alice will observe the interior mode $\overline{D}$ which is distinct from the original partner mode $\tilde{D}$ and is independent of initial states of the black hole. Hence, the monogamy of quantum entanglement is not violated. Recall that the approaches bundled under ``$A=R_{B}$'' try to interpret $\tilde{D}$ and $\overline{D}$ as the same physical degrees of freedom. Here we argued that they must be different. In fact, $\overline{D}$ and $\tilde{D}$ have no relevance at all as $\overline{D}$ is state-independent. Furthermore, $\overline{D}$ can be constructed in the same manner for young black holes~\footnote{
The AMPS argument relies on three key assumptions; a) unitarity b) no drama (i.e., smooth horizon) and c) effective quantum field theory outside the horizon. Note that a) results from quantum mechanics while b) and c) result from general relativity, namely the equivalence principle. We have assumed unitary quantum evolution and argued that the horizon must be smooth. Hence, the assumption c) must be violated. The essential point in our argument is that the outgoing mode $D$ is disentangled from the distilled qubit $\tilde{D}$. This conclusion was obtained from decay of OTOCs. Calculations of OTOCs generically go beyond n\"{a}ive effective quantum field theory. In its simplest form at the massless limit, the perturbation becomes a gravitational shockwave and leaves significant backreaction to the geometry. Such a calculation goes beyond effective quantum field theory on a fixed background geometry.}.  

It is worth comparing our proposal with the previous proposal by Maldacena and Susskind, an approach often called ``$\text{ER}=\text{EPR}$''. They argue that the distillation of the qubit $\tilde{D}$, as well as quantum operations afterwards by Bob, creates perturbations which will become a high-energy density near the horizon. Furthermore, they argue that this perturbation spoils the quantum entanglement between $D$ and $\tilde{D}$ and creates a firewall. As such, Alice is not able to observe entanglement between $D$ and $\tilde{D}$. 

While we also concluded that $D$ and $\tilde{D}$ are not entangled due to perturbations, the underlying physical mechanism differs in a crucial manner. In our scenario, $D$ and $\tilde{D}$ are disentangled due to the backreaction by the infalling observer herself, not due to Bob. Also, despite the loss of entanglement between $D$ and $\tilde{D}$, Alice still crosses the horizon smoothly and observe the entangled pair $D$ and $\overline{D}$ since $\overline{D}$ is realized as a mode different from the original mode $\tilde{D}$. After observing the interior mode $\overline{D}$, Alice may be (or may not be) hit by a gravitational shockwave which Bob creates, but this has nothing to do with quantum entanglement between $D$ and $\overline{D}$.

In summary, we propose the following resolution of the AMPS puzzle:

\begin{enumerate}[1)]
\item An infalling observer leaves non-trivial gravitational backreaction no matter how she falls. This disentangles the outgoing mode from the early radiation.
\item The infalling observer crosses the horizon smoothly and observes quantum entanglement between the outgoing mode and the interior mode.
\item The interior mode which the infalling observer sees is distinct from degrees of freedom that were originally entangled with the outgoing mode. Hence the monogamy of entanglement is not violated. 
\item Reconstructions of interior operators are not unique. Different reconstructions of the interior operators correspond to different infalling trajectories that the infalling observer follows.
\end{enumerate}

\begin{figure}
\centering
(a)\includegraphics[width=0.45\textwidth]{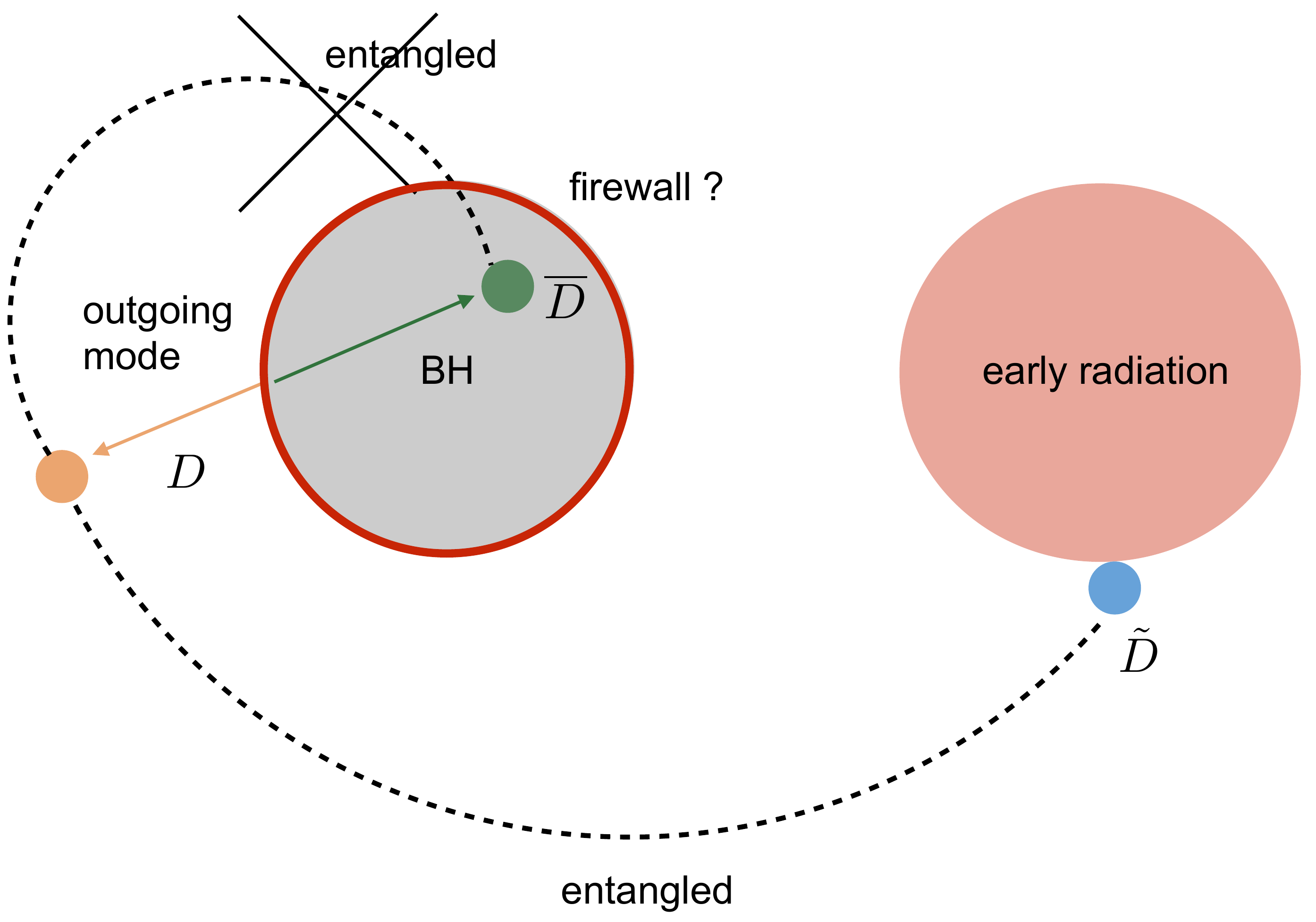} 
(b)\includegraphics[width=0.45\textwidth]{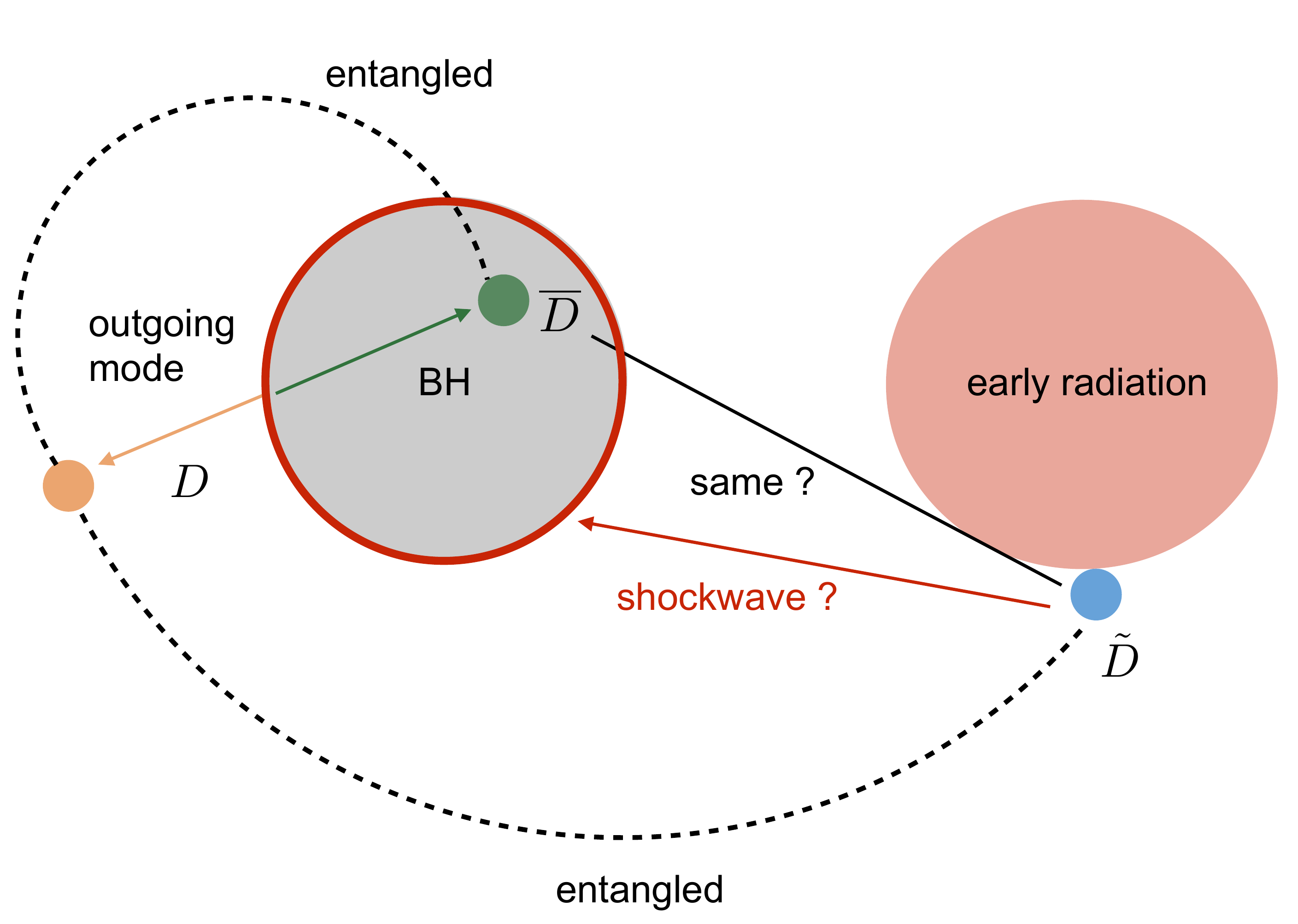}
(c)\includegraphics[width=0.45\textwidth]{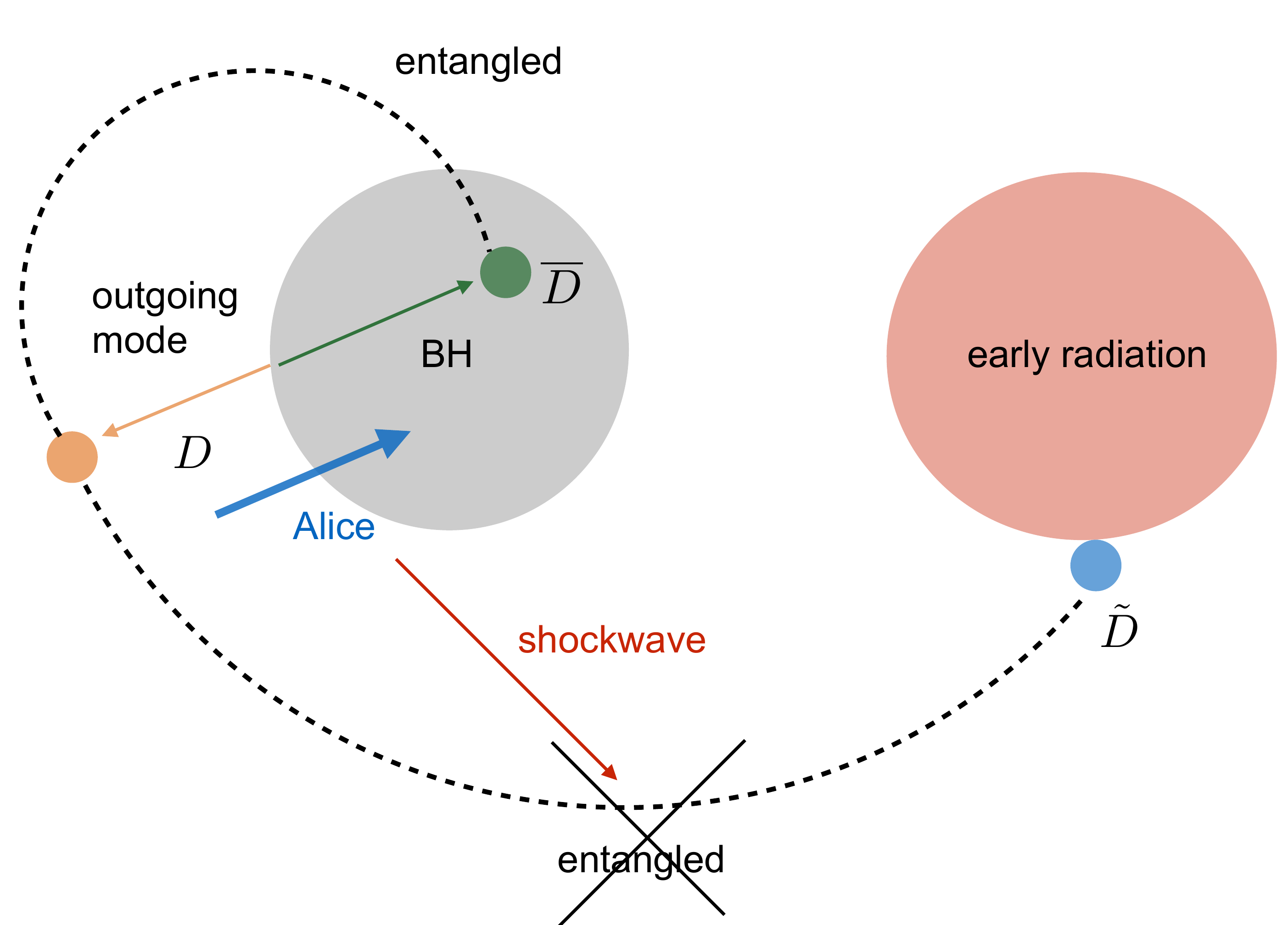}  
\caption{Schematic pictures of (a) the AMPS argument, (b) some previous proposals and (c) our proposal.
}
\label{fig-proposals}
\end{figure}

In Fig.~\ref{fig-proposals}, we summarized the original AMPS argument, previous proposals along the lines of $\text{ER}=\text{EPR}$, and our proposal.

\subsection{Revisiting firewalls}

We have argued that the infalling observer sees quantum entanglement while the outside observer does not due to the backreaction by the infalling observer. 
It is also intriguing to consider a possible scenario where the outside observer can see quantum entanglement while the infalling observer does not.
We have already refuted previous proposals which suggested that verification of quantum entanglement, namely the distillation of the entangled qubit $\tilde{D}$ on the early radiation $R$, generates a high-energy gravitational shockwave which prevents the infalling observer from crossing the horizon smoothly. 

Here we propose a physical mechanism for Bob to protect quantum entanglement from the backreaction by Alice. Recall that Alice can see quantum entanglement by interacting with earlier mode $A_{t}$. Bob's strategy, then, would be to retrieve $A_{t}$ from Alice and bring it back to his possession. One concrete (and peaceful) method is to run the Hayden-Preskill recovery protocol from the outside and pull Alice herself to the outside! That is, by preventing Alice from falling into a black hole, Bob is able to protect quantum entanglement between the outgoing mode and the early radiation. Since Alice does not cross the horizon, she will not see quantum entanglement. 

At this point, it is pretty satisfying to recall how the recovery protocols of~\cite{Yoshida:2017aa} work, reviewed in section~\ref{sec:HP}. The first protocol proceeds by verifying the entanglement between the outgoing mode $D$ and its partner $\overline{D}$. This is essentially Bob's attempt to observe the EPR pair between $D$ and $R$ by applying the EPR projection operator. The second protocol achieves this goal in a deterministic manner by unitarily restoring $D\overline{D}$ to be EPR pairs. Both protocols verify the EPR pair between $D$ and $R$, and steal the EPR pair from Alice by preventing her from crossing the horizon~\footnote{The first protocol projects $D\overline{D}$ onto EPR pair. Since $I(D,\overline{D})=\text{max}$, we have $I(D,AC)=0$ suggesting that $\overline{D}$ cannot be reconstructed on the left hand side. Since $I(A,\overline{A})\approx \text{max}$, Alice traverses the wormhole. So, we conclude that Alice does not see the EPR pair and simply returns to the outside. If we project $D\overline{D}$ onto other entangled states, such as $\frac{1}{\sqrt{2}} ( |01\rangle + |10 \rangle ) $, we still have $I(D,\overline{D})=\text{max}$ and $I(D,AC)=0$. However, $I(A,\overline{A})$ is not necessarily large. Hence, we conclude that Alice does not see the EPR pair. And unfortunately, she does not return to the outside.}. 

Based on this observation, we make the following proposal as physical interpretation of firewalls that may arise from the outside observer's verification of quantum entanglement:

\begin{enumerate}[1)]
\item An outside observer creates a non-trivial backreaction for an infalling observer by verifying (or protecting) quantum entanglement between the outgoing mode $D$ and the early radiation $R$.
\item However, it does not kill an infalling observer. Instead, it saves her from falling into a black hole recklessly. 
\item Since an infalling observer does not cross the horizon, she will not be able to see quantum entanglement. 
\end{enumerate}

The backreaction from Bob's verification of quantum entanglement is different from either high-energy barrier or some hard wall which would terminate the geometry at the horizon. It is worth noting that the Hayden-Preskill recovery, as well as the traversable wormhole, can be interpreted as a shockwave with negative energy. 

\subsection{Puzzle on non-local interactions (and its resolution)}

One mysterious feature of the AdS/CFT correspondence is an apparent non-locality behind the horizon in the two-sided black hole. The CFT Hamiltonian on the boundary is given by $H = H_{L} + H_{R}$ and the Hilbert space factorizes into the left and right. (See~\cite{Harlow:2016aa} for problems related to the factorization of the Hilbert space). Imagine two observers who start from left and right boundaries respectively and meet inside the black hole. This can be achieved in principle by starting at sufficiently early time. Two observers may interact with each other when they meet. From the perspective of the boundary quantum mechanical descriptions, however, this seems to imply some non-local coupling between left and right. 

We have encountered an analogous problem in the context of the AMPS puzzle where the infalling observer jumps into a black hole and sees the interior mode. Suppose that every possible representation of interior modes requires degrees of freedom on the right hand side. Suppose that she crosses the horizon smoothly and can observe the interior mode. This suggests that there must be some interaction between degrees of freedom on the left and right hand sides of the AdS black hole, leading to a contradiction. 

The resolution of this non-locality puzzle can be obtained straightforwardly. By falling into a black hole, Alice leaves a backreaction and creates a replicated mode $\overline{D}$ which consists exclusively of the left hand side degrees of freedom. Since she does not interact with $\tilde{D}$ on the right hand side, there is no contradiction~\footnote{A common misguided approach to the non-locality problem in the two-sided AdS black hole goes as follows. ``From the boundary perspective, whether Alice has interacted with the partner mode $\tilde{D}$ on the other side or not cannot be verified from either boundary because the interaction would happen behind the horizon. Yet, the interaction between Alice and $\tilde{D}$ can be seen by coupling two boundaries in a suitable manner, e.g. via the traversable wormhole effect. The interaction between Alice and $\tilde{D}$ does not lead to contradiction because two boundaries need to be coupled for verification of the interaction.'' We do not think this is a correct resolution of the non-locality problem. As we have rigorously demonstrated in this paper, the infalling observer will disentangle $D$ from $\tilde{D}$ and create a new interior mode $\overline{D}$ exclusively on her side of the boundaries. Namely, $\tilde{D}$ has nothing to do with her infalling experience. Furthermore, we think that it is possible to see the interior mode $\overline{D}$ via a careful quantum operation acting only on one side of the black hole, namely by performing the Hayden-Preskill recovery protocol on one side. See section~\ref{sec:peeking} for further discussions.}. Our viewpoint is that the black hole interior for an infalling observer emerges due to scrambling dynamics, not due to quantum entanglement between the left and right hand sides. We hope to address this issue further elsewhere.

\section{Discussions}\label{sec:discussion}

In this paper, we argued that the AMPS puzzle does not lead to inconsistencies by presenting state-independent reconstructions of interior operators. It is an interesting future problem to verify the proposed scenario in concrete models of quantum black holes. We are currently working on writing the interior operators in an explicit manner, namely in the SYK model. 

While we have focused on the AMPS puzzle in black hole horizons, it will be interesting to ask about cosmological horizons. In the dS space, a shockwave shifts the horizon in the opposite direction, making the Penrose diagram ``taller'', in a way similar to sending a negative energy shockwave in the AdS space. As such, our proposed resolution for the firewall problem in black hole horizons does not apply to the problem in cosmological horizons. One possible scenario may be that the backreaction by an existing observer recedes the horizon away so that she cannot actually cross the cosmological horizon.

In the reminder of the paper, we present discussions and speculations on relevant topics.

\subsection{Early time}

Since OTOCs start to decay significantly around the scrambling time, Bob can verify quantum entanglement between the outgoing mode $D$ and the early radiation $R$ when $\Delta t \ll t_{\text{scr}}$ even if Alice has fallen into a black hole. This, however, suggests that Alice should not be able to observe quantum entanglement when crossing the horizon! For this reason, our proposal resolves the AMPS puzzle only when $\Delta t \gtrapprox t_{\text{scr}}$. There are three possible physical explanations for the $D\overline{D}$ pair with $\Delta t \ll t_{\text{scr}}$. 

The first explanation concerns the applicability of effective quantum field theory near the singularity. In the Penrose diagram of the AdS black hole, one notices that Alice meets $\overline{D}$ very close to the singularity for $\Delta t \ll t_{\text{scr}}$. Hence the validity of quantum field theory is questionable. This observation seems to suggest that Alice may not be able to see quantum entanglement for small $\Delta t$. 

The second explanation concerns the quality of quantum entanglement that Alice may be able to see. Recall that the proper temperature near the Rindler horizon is given by $T = \frac{1}{2\pi \rho}$ where $\rho$ is the proper distance from the horizon. As one goes away from the horizon, the density of thermal entropy becomes smaller. Namely, one needs to coarse-grain a larger volume in order to distill a single EPR pair. 
In order for Alice to meet the outgoing mode $D$ near the horizon, she needs $\Delta t \gtrapprox r_{S} \log r_{S}$ in the Schwartzshild black hole which is of order the scrambling time. As such, we speculate that $\Delta t \gtrapprox r_{S} \log r_{S}$ is necessary for Alice to see a good quality EPR pair~\footnote{This observation enables us to obtain ``upper bound'' on the scrambling time. If the scrambling time $t_{\text{scr}}$, in a sense of decay of OTOCs, is longer than $\approx r_{S}\log r_{S}$, Bob can verify the EPR pair when $\Delta t \approx r_{S}\log r_{S}$. However, Alice can also see the EPR pair, leading to the violation of monogamy of entanglement. As such, we arrive at $t_{\text{scr}} \lessapprox r_{S}\log r_{S}$. Recalling that the Hayden-Preskill thought experiment says $t_{\text{scr}} \gtrapprox r_{S}\log r_{S}$, one may conclude $t_{\text{scr}}\approx r_{S}\log r_{S}$.}.

The third explanation is to assert that the loss of quantum entanglement for modes away from the horizon does not lead to significant violation of effective quantum field theory descriptions. 
The loss of entanglement in the Rindler modes near the horizon will create high energy density as entanglement is at short distance scale. 
For small $\Delta t$, however, the Rindler modes in question are separated by a long distance, and hence the loss of entanglement cannot be detected by simple local physical observables. Such tiny deviations may be superseded by e.g., curvature corrections. 

See Appendix~\ref{sec:CMT} for further discussions on the reconstruction of interior operators for small $\Delta t$. The important point is that construction of $\overline{D}$ is almost state-independent even when $\Delta t$ is small. We think that this phenomena may be of independent interest in the context of quantum many-body physics.

\subsection{Peeking in through the horizon}\label{sec:peeking}

\begin{figure}
\centering
\includegraphics[width=0.35\textwidth]{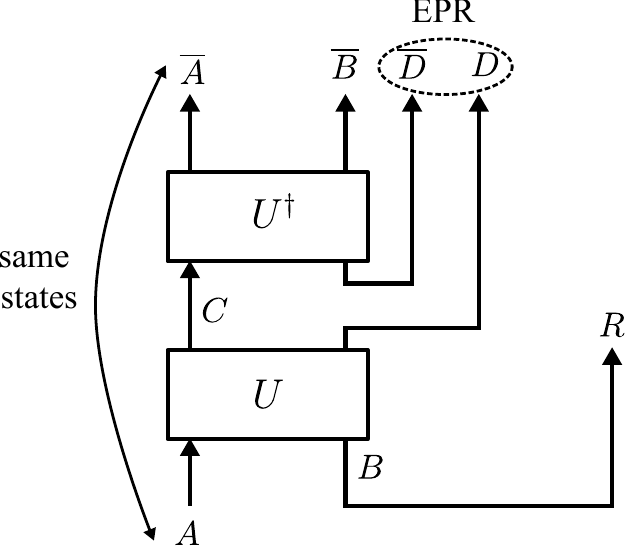} 
\caption{Reconstruction of the interior operator $\overline{D}$. If $D\overline{D}$ were to be measured in EPR pair, $A$ and $\overline{A}$ must be identical quantum states.
}
\label{fig-interior-reconstruction}
\end{figure}

While we have used the Hayden-Preskill recovery protocols as a mathematical tool to reconstruct interior operators, it is intriguing to think of physically implementing them on a black hole as an actual physical process. Let us focus on the probabilistic reconstruction protocol outlined in Section~\ref{sec:OTOC}. The initial state of the black hole is given by $|\text{EPR}\rangle_{A'A}|\text{EPR}\rangle_{BR}$ where $A'$ is the reference system of $A$. After the black hole evolves by a unitary operator $U$, the outgoing mode $D$ is kept outside the black hole. We prepare an additional EPR pair. The half of the EPR pair is thrown into a black hole while the other half, denoted by $\overline{D}$, is kept outside the black hole. We then implement the inverse $U^{\dagger}$ to the system. Finally, we project $\overline{A'}\overline{A}$ onto the EPR pair. This leaves $\overline{D}D$ to be a nearly ideal EPR pair. Fig.~\ref{fig-interior-reconstruction} summarizes the reconstruction of the interior operator as a physical process. 

Let us ponder over possible geometric interpretation of these protocols. Since $\overline{A'}\overline{A}$ is projected to the EPR pair, an input quantum state on $A$ would remain invariant and reappear as the same quantum state on $\overline{A}$. This suggests that the infalling observer, who fell into a black hole, returns to the outside again safely~\footnote{More precisely, we applied a projection operator to pick an event where the infalling observer returns safely. This works only probabilistically. One may apply the deterministic version of the Hayden-Preskill recovery protocol to make this happen deterministically.} This is not surprising as we applied the inverse $U^{\dagger}$. What is surprising is that the outgoing mode $D$ is entangled with another mode $\overline{D}$ which is left outside the black hole. One possible interpretation is that, due to the backreaction of the Hayden-Preskill recovery protocol, the interior operator $\overline{D}$ is somehow pulled to the outside of the black hole. Hence, the infalling observer appears to cross the horizon smoothly and returns safely with the interior mode in hand!

As this speculation suggests, reconstruction of the interior operators by the Hayden-Preskill recovery protocols enables us to probe the physics behind the horizon from the outside even when the black hole is one-sided. To turn the speculation into a concrete observation, it may be interesting to consider a similar process by using the Gao-Jafferis-Wall traversable wormhole. 

\subsection{Entanglement wedge reconstruction}

We have presented a protocol to reconstruct interior operators by using the Hayden-Preskill recovery. Hence it is reasonable to expect that the Hayden-Preskill recovery is useful in reconstructing operators in the entanglement wedge. Progresses along this line have been recently made by Almheiri~\cite{Almheiri2018}. Since operators in the entanglement wedge must be interpreted as quantum field theory operators on a fixed geometry, state-independent reconstructions must be possible. We expect that our results are useful in achieving this goal.

\section*{Acknowledgment}

I would like to thank Ahmed Almheiri, Yoni BenTov, Isaac Kim, Rob Myers and Herman Verlinde for useful discussions on related topics. I thank Yoni BenTov and Rob Myers for careful reading of the preliminary draft and a number of useful feedbacks. For the second version of the paper, I would like to thank Ning Bao, Raphael Bousso, Juan Maldacena, Xiaoliang Qi, Suvrat Raju, Lenny Susskind and Ying Zhao for useful discussions. Acknowledgment of discussions with a person does not necessarily imply the person's agreement. Research at the Perimeter Institute is supported by the Government of Canada through Innovation, Science and Economic Development Canada and by the Province of Ontario through the Ministry of Economic Development, Job Creation and Trade.

\appendix

\section{Energy measurement}\label{sec:interaction}

We have presented a protocol to reconstruct the interior operator $\overline{D}$ without involving the early radiation $R$. Our protocol requires EPR pairs in the ancilla system $E\overline{E}$ which interact with the earlier mode $A$ in a particular manner. Here we argue that much simpler and generic interaction can achieve this goal.

To set the stage, we elaborate on the issue briefly mentioned in Section~\ref{sec:AMPS_outside}. Let us denote the orthonormal basis of the outgoing mode $D$ by $|n\rangle \in \mathcal{H}_{D}$. According to the standard rules of statistical physics, the reduced density operator on $D$ should look thermal:
\begin{align}
\rho_{D} = \sum_{n} w_{n} |n\rangle \langle n|, \qquad w_{n} = \frac{e^{-\beta E_{n}}}{Z}
\end{align}
with $Z = \sum_{n} e^{-\beta E_{n}}$. The discrepancy with $\rho_{D}$ in Eq.~\eqref{eq:no_correlation} originates from the total energy conservation. In particular, energy densities on $C$ an $D$ are correlated. An analysis with energy conservation taken into consideration indicates that $\rho_{CD}$ sustains diagonal correlations in the $|n\rangle$-basis while off-diagonal correlations decohere due to scrambling dynamics (i.e., correlations are classical)~\cite{Beni18}. 

The above observation prompts us to consider two different classes of operators on $D$. The off-diagonal operators are the ones which may change the energy eigenstates (or the particle number) in $D$. The interior partners of those can be identified in $C$ as operators which decrease and/or increase the energy density. This is due to diagonal correlations in $\rho_{CD}$. The diagonal operators are the ones which leave energy eigenstates unchanged up to phases. Such operators can be explicitly written as follows:
\begin{align}
O_{\theta} = \sum_{n}e^{i \theta_{n}} |n\rangle\langle n|.
\end{align}
Here $\theta$ represents all the data about $\theta_{n}$. Unlike off-diagonal operators, the interior partners of diagonal operators $O_{\theta}$ cannot be found in $C$~\footnote{Off-diagonal partners exist due to the existence of diagonal correlations. Diagonal partners do not exist due to the absence of off-diagonal correlations.}. 

Off-diagonal correlations can be restored by simply measuring the energy of the earlier mode $A$. The procedure of quantum measurement can be formulated as a unitary process in the following manner. We adjoin the system to a pointer system $E$ whose dimensionality is equal to $A$. Denote the orthonormal basis by $|n\rangle_{E}$. If we start the system in an arbitrary pure state $|\psi \rangle = \sum_{n}c_{n} |n\rangle_{A}$, the measurement leads to the evolution:
\begin{align}
|\psi\rangle_{A} |0\rangle_{E} \rightarrow \sum_{n} c_{n} |n\rangle_{A}|n\rangle_{E}
\end{align}
under the unitary operator
\begin{align}
W = \sum_{n} |n\rangle \langle n|_{A} \otimes {X_{E}}^n
\end{align}
where $X= \sum_{j} | j+1\rangle \langle j |$ is the Weyl operator (a generalization of Pauli-$X$ to multi-level systems).

The important point here is that applying $M$ allows Alice to ``copy'' $O_{\theta}$ from $A$ to $E$. This can be most explicitly shown by noting the following identity:
\begin{align}
\figbox{1.0}{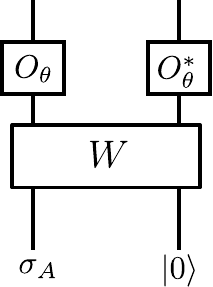} \ = \ \figbox{1.0}{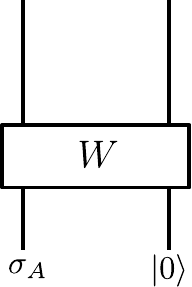}
\end{align}
which holds for any density operator $\sigma_{A}$. Hence, from the outside description, all Alice needs to do is measure the energy (or the particle number) of the earlier mode $A$. The interior operator $\overline{D}$ can be reconstructed in the pointer $E$ and the remaining black hole $C$. 

\section{Verlinde-Verlinde proposal}\label{sec:Verlinde}

Verlinde and Verlinde proposed an intriguing resolution of the AMPS puzzle by employing the idea of quantum error-correction~\cite{Verlinde:2013aa}. To illustrate the idea in a simplified setting, suppose that a black hole $B$ is entangled with the early radiation $R$ only in the codeword subspace $\mathcal{H}_{B_{0}}\subset \mathcal{H}_{B}$ with $\mathcal{H}_{B_{0}} = \{ |\tilde{j}\rangle \}_{j=1}^{|B_{0}|}$. Specifically, consider a quantum state given by
\begin{align}
|\Psi \rangle_{BR} = \frac{1}{\sqrt{|B_{0}|}} \sum_{j=1}^{|B_{0}|} |\tilde{j}\rangle_{B}\otimes |\tilde{j}\rangle_{R}\label{eq:code}
\end{align}
where $|B_{0}| \ll |B|$. Here, by the ``codeword subspace'', we simply mean that $|B_{0}|<|B|$. If the black hole evolves by a Haar random unitary, the mutual information $I(C,D) $ is nearly maximal, implying that $\overline{D}$ can be reconstructed on $C$. Indeed, they performed a careful statistical physics analysis and presented explicit constructions of interior operators by using the error recovery procedure known in quantum error-correction theory. A related idea was considered~\cite{Bao:2018aa} in a slightly different context.  

In our interpretation, however, this observation itself does not resolve the AMPS puzzle. First, this proposal does not apply to black holes which have maximal entanglement such as the eternal AdS black hole. Let $S_{\text{BH}}$ be the coarse-grained entropy of the black hole and $S_{\text{ent}}$ be the entanglement entropy between the black hole and the radiation $R$. Then the black holes under consideration satisfy $S_{\text{BH}} >  S_{\text{ent}}$ whereas the eternal AdS black hole corresponds to the case with $S_{\text{BH}} =  S_{\text{ent}}$. The situation of Eq.~\eqref{eq:code} resembles black holes before the Page time. If such a codeword subspace emerges, its origin should be explained. Second, the proposed reconstruction of interior operators depends on choices of the codeword subspace, and as such, is state-dependent. Third, as pointed out in~\cite{Beni18}, this proposal runs into another problem where the Hayden-Preskill recovery cannot be performed. Indeed, if the outgoing mode $D$ can be used for the Hayden-Preskill recovery, then it should not be entangled with $C$. 

Nevertheless, we will make an interesting link between our proposal and their viewpoint. Let us view the energy measurement discussed in the appendix~\ref{sec:interaction} as a process of ``collapsing'' $A$ into some pure state without introducing the pointer system. If a particular energy eigenstate $|n\rangle$ is measured, the postselected black hole dynamics can be interpreted as an isometry from $B$ to $CD$:
\begin{align}
U |0\rangle_{A} : \mathcal{H}_{B} \rightarrow \mathcal{H}_{C}\otimes \mathcal{H}_{D} \qquad |B| < |C||D|.
\end{align}
An alternative interpretation of this dynamics is to think that the black hole is represented by $S_{\text{BH}}=\log_{2} |A||B|$ qubits, but its entanglement entropy is $S_{\text{ent}} = \log_{2} |B|$. In other words, measurement on $A$ reduced the entanglement entropy of the black hole. Here $B$ is viewed as the subspace where the black hole is entangled with the partner. This situation resembles the scenario considered by Verlinde and Verlinde~\cite{Verlinde:2013aa}. By having $S_{\text{ent}} < S_{\text{BH}}$, quantum entanglement can be restored between $C$ and $D$. We have treated a similar situation in section~\ref{sec:outside} by including a measurement apparatus $M$.

\begin{figure}
\centering
\includegraphics[width=0.14\textwidth]{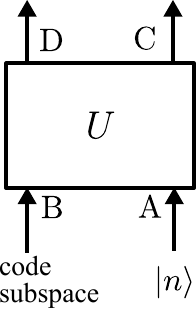} 
\caption{A black hole entangled through a codeword subspace. An infalling observer measures $|n\rangle$ on $A$, and quantum entanglement between $C$ and $D$ is restored.
}
\label{fig-subspace}
\end{figure}

In fact, collapsing the input Hilbert space by generic projective measurements on $A$ restores quantum entanglement between $C$ and $D$. For simplicity of discussion, let us assume that $U$ is Haar random. An explicit calculation based on Haar random unitary suggests that projective measurements on A with $|A| \gtrapprox |D|^2$ leads to nearly maximal mutual information $I(C,D)$~\cite{Beni18}. When thermal correlations are taken into account, only the off-diagonal correlations need to be restored, so taking $|A|\gtrapprox |D|$ suffices. The important implication of this observation is that (essentially) any measurement by an infalling observer effectively creates a situation where the black hole is entangled through a codeword subspace. This restores quantum entanglement in $C$ and $D$, and does not require any fine-tuning. 

The work by Verlinde and Verlinde did not provide concrete account for the physical origin of the codeword subspace. We think that inclusion of the infalling observer effectively creates situations similar to theirs.

\section{Fault-tolerant entanglement}\label{sec:CMT}

The state-independent reconstruction of interior operators may be of independent interest from the perspective of quantum many-body physics as it reveals a certain universal aspect of quantum entanglement in chaotic quantum dynamics. 

Let us consider an initial quantum state of the following form:
\begin{align}
\rho_{\text{in}} = |\psi\rangle \langle \psi|_{A} \otimes \sigma_{B}
\end{align}
where $|\psi\rangle$ is a fixed pure state on $A$ and $\sigma_{B}$ is an arbitrary quantum state, pure or mixed, on $B$. The system evolves under a chaotic unitary dynamics $U$:
\begin{align}
\rho_{\text{out}} = U\rho_{\text{in}}U^{\dagger}. 
\end{align}
Here we make two assumptions. First, we assume that OTOCs for $O_{D}(t)$ and $O_{A}=|\psi\rangle\langle \psi|$ decay to small values. Second, we assume that the system thermalizes, in a sense that time-ordered correlation functions, such as $\langle O_{D}(t) O_{A}(0) \rangle$, decay to small values. From these assumptions, we can find that $D$ is nearly maximally entangled with its complement $C$ as long as $|A| \gtrapprox |D|^2$. Here our primary interest is in how $C$ and $D$ are entangled.

In this paper, we argued that the partner operators can be constructed on $C$ in a way independent of $\sigma_{B}$. This implies that $D$ is entangled with some fixed degrees of freedom, denoted by $\overline{D}$ inside $C$, regardless of the choice of the initial state $\sigma_{B}$. Here $\overline{D}$ may be interpreted as a code subspace where entanglement with $D$ is fault-tolerantly stored in such a way that is protected from any perturbations on $\sigma_{B}$. In other words, this subspace is determined by the dynamics $U$ and $|\psi\rangle_{A}$, but is independent of $\sigma_{B}$. As such, the entanglement structure between $C$ and $D$ is a universal feature of the chaotic dynamics $U$. 

The above observation assumed that OTOCs have decayed to small values. Next, let us consider situations where OTOCs have not decayed to smaller values yet, e.g., before the scrambling time. Given an operator $O_{D}$ on $D$, let us reconstruct a partner operator $O_{\overline{D}}$ supported on $C$ by using the Hayden-Preskill recovery protocol. Consider the maximally mixed state $\sigma_{B} = \frac{1}{d_{B}}I_{B}$ as the initial state. Since $C$ and $D$ are entangled only weakly, we will have 
\begin{align}
\alpha \equiv \langle O_{D} \otimes O_{\overline{D}} \rangle_{\sigma_{B} = \frac{1}{d_{B}}I_{B}} \ll 1
\end{align}
where the expectation value is taken with respect to $\rho_{\text{out}}$. Note that the expectation value would be close to $1$ if $C$ and $D$ were nearly maximally entangled. 

Despite the fact that $O_{D}$ and $O_{\overline{D}}$ are only weakly correlated, the construction of $O_{\overline{D}}$ is (almost) state-independent in the following sense. To be specific, let us consider pure states $\sigma_{B}=|\phi\rangle \langle \phi|$ on $B$ as initial states. We then typically have 
\begin{align}
\int d | \phi\rangle \ \langle O_{D} \otimes O_{\overline{D}} \rangle_{\sigma_{B}=|\phi\rangle\langle \phi|} = \alpha
\end{align}
where the integral is taken uniformly over all $|\phi\rangle$ on $B$. Hence, if $|\phi\rangle$ is a typical pure state sampled from the Haar measure (or an ensemble forming $2$-design), then we have 
\begin{align}
\langle O_{D} \otimes O_{\overline{D}} \rangle_{\sigma_{B}=|\phi\rangle\langle \phi|} \approx \alpha
\end{align}
where the deviation from $\alpha$ is exponentially suppressed. As such, the structure of entanglement between $D$ and $\overline{D}$ is universal even when OTOCs have not decayed to small values yet.

\section{Alice's quantum field theory}\label{sec:Alice}

We discuss effective quantum field theory that the infalling observer may experience. According to the equivalence principle, we expect that Alice will have normal quantum mechanical descriptions until she reaches the singularity. The fundamental difficulty in addressing this question lies in the very fact that Alice never returns to the outside. To circumvent this, we may think of performing the Hayden-Preskill recovery protocol from the outside so that Alice can have effective quantum field theory without reaching the singularity. To understand quantum field theory under the effect of the Hayden-Preskill recovery, we must understand its effect on the black hole geometry. At this moment, we are unsure about what is the corresponding geometric picture. It is even unclear to us if it admits any classical geometric description or not. Below we present some speculation on the backreacted geometry.

\begin{figure}
\centering
(a)\includegraphics[width=0.35\textwidth]{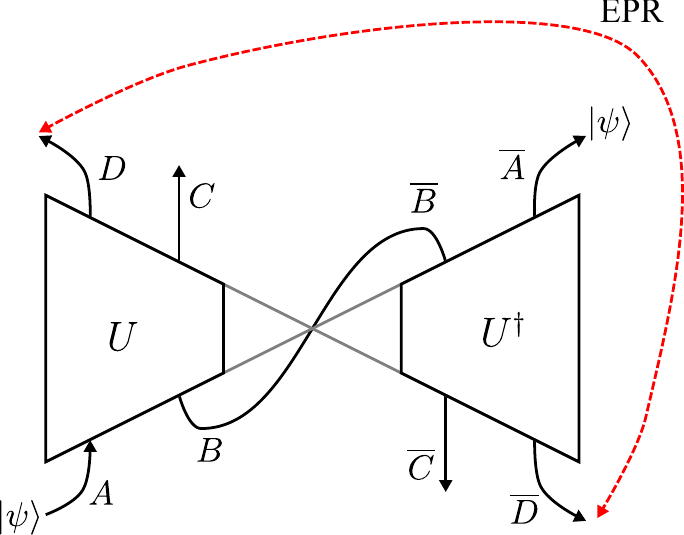} 
(b)\includegraphics[width=0.35\textwidth]{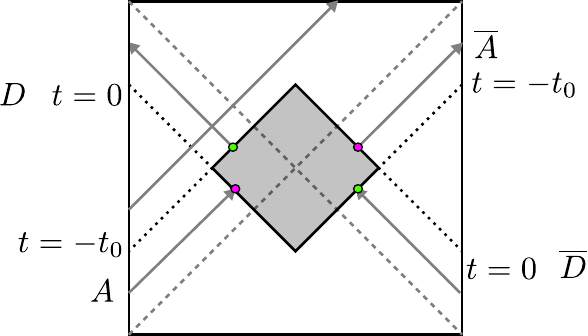} 
(c)\includegraphics[width=0.25\textwidth]{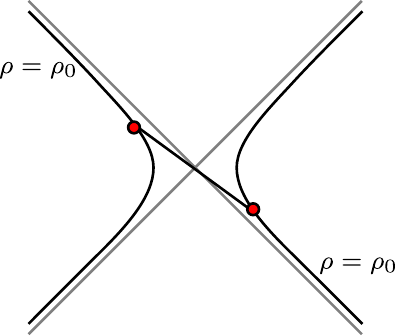} 
\caption{(a) The Hayden-Preskill recovery protocol. (b) Backreaction from the Hayden-Preskill recovery. (c) Alice's quantum field theory near the horizon. Two surfaces in the opposite regions are identified. 
}
\label{fig-backreaction}
\end{figure}

Let us focus on the first recovery protocol from~\cite{Yoshida:2017aa} which projects $D\overline{D}$ into the EPR pair. To gain some insights, it is useful to draw the process in the Penrose diagram of the eternal two-sided AdS black hole. Let us take the time directions to be upward and downward on left and right hand sides respectively. One may flip the diagram of the recovery procedure in accordance with the convention of the Penrose diagram. The flipped diagram is shown in Fig.~\ref{fig-backreaction}(a). This cartoon picture enables us to depict the recovery protocol schematically in the Penrose diagram as in Fig.~\ref{fig-backreaction}(b). Here two modes $D\overline{D}$ on two sides are collapsed into EPR pairs. This lets $A$ traverse across the black hole and reach the other boundary at $\overline{A}$. The backreaction by the Hayden-Preskill recovery becomes significant when the separation of the time between the infalling observer $A$ and the outgoing mode $D$ becomes of order the scrambling time. In the Penrose diagram, let us take $D$ and $\overline{D}$ at $t=0$ on the left and right hand sides respectively. The geometry will be significantly modified only inside the shaded diamond shown in Fig.~\ref{fig-backreaction}(b) which is constructed by considering modes at $t = - t_{0}$ with $t_{0}\approx t_{\text{scr}}$. If a signal is sent between $t = - t_{0}$ and $t=0$, it will cross the horizon and reach the singularity. If a signal is sent before $t= - t_{0}$, it will traverse to the other side. These observations suggest that the portion inside the diamond effectively disappears, and antipodal points should be identified as shown in Fig.~\ref{fig-backreaction}(b) so that signals entering the diamond can traverse. Precise value of $t_{0}$ will depend on the size of $D$ as well as the amount of quantum information in $A$ we wish to transmit
~\footnote{For simplicity of discussion, consider a deeply scrambling regime with $t_{0} \gtrapprox t_{\text{scr}}$. An analysis from~\cite{Yoshida:2017aa} suggests that $|D|\gtrapprox |A|$ suffices to transmit all the quantum information contained inside $A$. If quantum information in the subspace $A_{0}\subset A$ are to be transmitted, we only need $|D|^2 \gtrapprox |A||A_{0}|$. So, even when $A$ is a big subsystem, $|D|^2 \gtrapprox \text{const}\cdot |A|$ suffices if $A_{0}$ is small. In a qubit count, the sufficient condition would be $n_{D}\gtrapprox \frac{1}{2}n_{A}$.}. 

Suppose that Alice jumps into a black hole at $t < - t_{0}$. She will enter the shaded region near the horizon which is approximated by the Rindler space. Letting $\rho$ be the distance from the horizon, Alice sees the backreaction from the Hayden-Preskill recovery at the surface $\rho=\rho_{0}$. The recovery protocol prevents Alice from seeing EPR pairs at $\rho \leq \rho_{0}$ and lets her traverse to the opposite region. This situation can be realized by cutting the Rindler space at $\rho = \rho_{0}$ and glue two regions together. If the time separation between $A$ and $D$ is of order the scrambling time, $\rho_{0}$ is small compared to the black hole radius, but is still large compared to the Planck length. We speculate that Alice's effective quantum field theory is defined on this truncated and traversable Rindler space. If the separation becomes larger than the scrambling time, $\rho_{0}$ eventually becomes of order the Planck length. In this regime, we expect that the effect of the Hayden-Preskill recovery can be simply treated as identification of opposite sides by neglecting physics below the Planck scale. 

\providecommand{\href}[2]{#2}\begingroup\raggedright\endgroup


\begin{thebibliography}{10}

\bibitem{Hawking:1976aa}
S.~W. Hawking, ``Breakdown of predictability in gravitational collapse,''
  \href{http://dx.doi.org/10.1103/PhysRevD.14.2460}{{\em Phys. Rev. D}
  {\bfseries 14} (1976) 2460--2473}.

\bibitem{Almheiri13}
A.~Almheiri, D.~Marolf, J.~Polchinski, and J.~Sully, ``Black holes:
  complementarity or firewalls?,''
  \href{http://dx.doi.org/10.1007/JHEP02(2013)062}{{\em JHEP} {\bfseries 02}
  (2013) 062}.

\bibitem{Hawking75}
S.~Hawking, ``Particle creation by black holes,'' {\em Commun. Math. Phys.}
  {\bfseries 43} (1975) 199--220.

\bibitem{Almheiri13b}
A.~Almheiri, D.~Marolf, J.~Polchinski, D.~Stanford, and J.~Sully, ``An apologia
  for firewalls,'' \href{http://dx.doi.org/10.1007/JHEP09(2013)018}{{\em JHEP}
  {\bfseries 09} (2013) 018}.

\bibitem{Papadodimas:2013aa}
K.~Papadodimas and S.~Raju, ``An infalling observer in ads/cft,''
  \href{http://dx.doi.org/10.1007/JHEP10(2013)212}{{\em JHEP} {\bfseries 10}
  (2013) 212}.

\bibitem{Maldacena13}
J.~Maldacena and L.~Susskind, ``Cool horizons for entangled black holes,''
  \href{http://dx.doi.org/10.1002/prop.201300020}{{\em Fortsch. Phys.}
  {\bfseries 61} (2013) 781--811}.

\bibitem{Susskind13}
L.~Susskind, ``New concepts for old black holes.'' Arxiv:1311.3335.

\bibitem{Bousso:2013ab}
R.~Bousso, ``Complementarity is not enough,''
  \href{http://dx.doi.org/10.1103/PhysRevD.87.124023}{{\em Phys. Rev. D}
  {\bfseries 87} (2013) 124023--}.

\bibitem{Marolf:2013aa}
D.~Marolf and J.~Polchinski, ``Gauge-gravity duality and the black hole
  interior,'' \href{http://dx.doi.org/10.1103/PhysRevLett.111.171301}{{\em
  Phys. Rev. Lett.} {\bfseries 111} (2013) 171301--}.

\bibitem{Bousso:2013aa}
R.~Bousso, ``Firewalls from double purity,''
  \href{http://dx.doi.org/10.1103/PhysRevD.88.084035}{{\em Phys. Rev. D}
  {\bfseries 88} (2013) 084035--}.

\bibitem{Giddings:2013aa}
S.~B. Giddings, ``Nonviolent nonlocality,''
  \href{http://dx.doi.org/10.1103/PhysRevD.88.064023}{{\em Phys. Rev. D}
  {\bfseries 88} (2013) 064023--}.

\bibitem{Harlow:2016ab}
D.~Harlow, ``Jerusalem lectures on black holes and quantum information,''
  \href{http://dx.doi.org/10.1103/RevModPhys.88.015002}{{\em Rev. Mod. Phys.}
  {\bfseries 88} (2016) 015002--}.

\bibitem{Bousso:2014aa}
R.~Bousso, ``Violations of the equivalence principle by a nonlocally
  reconstructed vacuum at the black hole horizon,''
  \href{http://dx.doi.org/10.1103/PhysRevLett.112.041102}{{\em Phys. Rev.
  Lett.} {\bfseries 112} (2014) 041102--}.

\bibitem{Marolf:2016aa}
D.~Marolf and J.~Polchinski, ``Violations of the born rule in cool
  state-dependent horizons,''
  \href{http://dx.doi.org/10.1007/JHEP01(2016)008}{{\em JHEP} {\bfseries 1}
  (2016) 8}.

\bibitem{Papadodimas:2014aa}
K.~Papadodimas and S.~Raju, ``Black hole interior in the holographic
  correspondence and the information paradox,''
  \href{http://dx.doi.org/10.1103/PhysRevLett.112.051301}{{\em Phys. Rev.
  Lett.} {\bfseries 112} (2014) 051301--}.

\bibitem{Papadodimas:2014ab}
K.~Papadodimas and S.~Raju, ``State-dependent bulk-boundary maps and black hole
  complementarity,'' \href{http://dx.doi.org/10.1103/PhysRevD.89.086010}{{\em
  Phys. Rev. D} {\bfseries 89} (2014) 086010--}.

\bibitem{Kourkoulou17}
I.~Kourkoulou and J.~Maldacena, ``Pure states in the syk model and nearly-ads2
  gravity.'' Arxiv:1707.02325.

\bibitem{Almheiri2018}
A.~Almheiri, ``Holographic quantum error correction and the projected black
  hole interior,'' \href{http://arxiv.org/abs/arXiv:1810.02055}{{\ttfamily
  arXiv:1810.02055}}.

\bibitem{Harlow:2014aa}
D.~Harlow, ``Aspects of the papadodimas-raju proposal for the black hole
  interior,'' \href{http://dx.doi.org/10.1007/JHEP11(2014)055}{{\em JHEP}
  {\bfseries 11} (2014) 55}.

\bibitem{Almheiri:2015ac}
A.~Almheiri, X.~Dong, and D.~Harlow, ``Bulk locality and quantum error
  correction in ads/cft,''
  \href{http://dx.doi.org/10.1007/JHEP04(2015)163}{{\em JHEP} {\bfseries 4}
  (2015) 163}.

\bibitem{Pastawski15b}
F.~Pastawski, B.~Yoshida, D.~Harlow, and J.~Preskill, ``Holographic quantum
  error-correcting codes: toy models for the bulk/boundary correspondence,''
  \href{http://dx.doi.org/10.1007/JHEP06(2015)149}{{\em JHEP} {\bfseries 06}
  (2015) 149}.

\bibitem{Verlinde:2013aa}
E.~Verlinde and H.~Verlinde, ``Black hole entanglement and quantum error
  correction,'' \href{http://dx.doi.org/10.1007/JHEP10(2013)107}{{\em JHEP}
  {\bfseries 10} (2013) 107}.

\bibitem{Verlinde13a}
E.~Verlinde and H.~Verlinde, ``Behind the horizon in ads/cft.''
  Arxiv:1311.1137.

\bibitem{Verlinde13b}
E.~Verlinde and H.~Verlinde, ``Passing through the firewall.'' Arxiv:1306.0515.

\bibitem{Goel18}
A.~Goel, H.~T. Lam, G.~J. Turiaci, and H.~Verlinde, ``Expanding the black hole
  interior: Partially entangled thermal states in syk.'' Arxiv:1807.03916.

\bibitem{Beni18}
B.~Yoshida, ``Soft mode and interior operator in hayden-preskill thought
  experiment.'' Arxiv:1812.07353.

\bibitem{Hayden07}
P.~Hayden and J.~Preskill, ``Black holes as mirrors: quantum information in
  random subsystems,'' {\em JHEP} {\bfseries 09} (2007) 120.

\bibitem{Hosur:2015ylk}
P.~Hosur, X.-L. Qi, D.~A. Roberts, and B.~Yoshida, ``{Chaos in quantum
  channels},'' \href{http://dx.doi.org/10.1007/JHEP02(2016)004}{{\em JHEP}
  {\bfseries 02} (2016) 004}.

\bibitem{Maldacena:2016aa}
J.~Maldacena, S.~H. Shenker, and D.~Stanford, ``A bound on chaos,''
  \href{http://dx.doi.org/10.1007/JHEP08(2016)106}{{\em JHEP} {\bfseries 8}
  (2016) 106}.

\bibitem{Harlow:2013aa}
D.~Harlow and P.~Hayden, ``Quantum computation vs. firewalls,''
  \href{http://dx.doi.org/10.1007/JHEP06(2013)085}{{\em JHEP} {\bfseries 6}
  (2013) 85}.

\bibitem{Yoshida:2017aa}
B.~Yoshida and A.~Kitaev, ``Efficient decoding for the hayden-preskill
  protocol,'' \href{http://arxiv.org/abs/arXiv:1710.03363}{{\ttfamily
  arXiv:1710.03363}}.

\bibitem{Page93}
D.~N. Page, ``Average entropy of a subsystem,'' {\em Phys. Rev. Lett.}
  {\bfseries 71} (1993) 1291--1294.

\bibitem{Page:1993aa}
D.~N. Page, ``Information in black hole radiation,''
  \href{http://dx.doi.org/10.1103/PhysRevLett.71.3743}{{\em Phys. Rev. Lett.}
  {\bfseries 71} (1993) 3743--3746}.

\bibitem{Mathur:2009aa}
S.~D. Mathur, ``The information paradox: a pedagogical introduction,''
  \href{http://dx.doi.org/10.1088/0264-9381/26/22/224001}{{\em Class. Quant.
  Grav.} {\bfseries 26} (2009) 224001}.

\bibitem{Braunstein:2013aa}
S.~L. Braunstein, S.~Pirandola, and K.~{\.Z}yczkowski, ``Better late than
  never: Information retrieval from black holes,''
  \href{http://dx.doi.org/10.1103/PhysRevLett.110.101301}{{\em Phys. Rev. Lett}
  {\bfseries 110} (2013) 101301}.

\bibitem{Susskind18}
L.~Susskind, ``Black holes and complexity classes.'' Arxiv:1802.02175.

\bibitem{Aaronson16}
S.~Aaronson, ``The complexity of quantum states and transformations: From
  quantum money to black holes.'' Arxiv:1607.05256.

\bibitem{Oppenheim:2014aa}
J.~Oppenheim and B.~Unruh, ``Firewalls and flat mirrors: An alternative to the
  amps experiment which evades the harlow-hayden obstacle,''
  \href{http://dx.doi.org/10.1007/JHEP03(2014)120}{{\em JHEP} {\bfseries 3}
  (2014) 120}.

\bibitem{t-Hooft:1990aa}
G.~'t~Hooft, ``The black hole interpretation of string theory,''
  \href{http://dx.doi.org/https://doi.org/10.1016/0550-3213(90)90174-C}{{\em
  Nucl. Phys. B} {\bfseries 335} (1990) 138--154}.

\bibitem{t-Hooft:1985aa}
G.~'t~Hooft, ``On the quantum structure of a black hole,''
  \href{http://dx.doi.org/https://doi.org/10.1016/0550-3213(85)90418-3}{{\em
  Nucl. Phys. B} {\bfseries 256} (1985) 727--745}.

\bibitem{Susskind:1994aa}
L.~Susskind and L.~Thorlacius, ``Gedanken experiments involving black holes,''
  \href{http://dx.doi.org/10.1103/PhysRevD.49.966}{{\em Phys. Rev. D}
  {\bfseries 49} (1994) 966--974}.

\bibitem{Susskind:1993aa}
L.~Susskind, L.~Thorlacius, and J.~Uglum, ``The stretched horizon and black
  hole complementarity,''
  \href{http://dx.doi.org/10.1103/PhysRevD.48.3743}{{\em Phys. Rev. D}
  {\bfseries 48} no.~8, (10, 1993) 3743--3761}.

\bibitem{Lashkari13}
N.~Lashkari, D.~Stanford, M.~Hastings, T.~Osborne, and P.~Hayden, ``Towards the
  fast scrambling conjecture,''
  \href{http://dx.doi.org/10.1007/JHEP04(2013)022}{{\em JHEP} {\bfseries 04}
  (2013) 22}.

\bibitem{Kitaev_unpublished}
A.~Kitaev. Unpublished.

\bibitem{Almheiri:2015ab}
A.~Almheiri and J.~Polchinski, ``Models of ads2 backreaction and holography,''
  \href{http://dx.doi.org/10.1007/JHEP11(2015)014}{{\em JHEP} {\bfseries 11}
  (2015) 14}.

\bibitem{Maldacena:2016ab}
J.~Maldacena and D.~Stanford, ``Remarks on the sachdev-ye-kitaev model,''
  \href{http://dx.doi.org/10.1103/PhysRevD.94.106002}{{\em Phys. Rev. D}
  {\bfseries 94} (2016) 106002--}.

\bibitem{Engelsoy:2016aa}
J.~Engels{\"o}y, T.~G. Mertens, and H.~Verlinde, ``An investigation of ads2
  backreaction and holography,''
  \href{http://dx.doi.org/10.1007/JHEP07(2016)139}{{\em JHEP} {\bfseries 7}
  (2016) 139}.

\bibitem{Stanford:2016aa}
D.~Stanford, J.~Maldacena, and Z.~Yang, ``Conformal symmetry and its breaking
  in two-dimensional nearly anti-de sitter space,'' {\em PTEP} (2016) 12C104.

\bibitem{Gao:2017aa}
P.~Gao, D.~L. Jafferis, and A.~C. Wall, ``Traversable wormholes via a double
  trace deformation,'' \href{http://dx.doi.org/10.1007/JHEP12(2017)151}{{\em
  JHEP} {\bfseries 12} (2017) 151}.

\bibitem{Traversable2017}
J.~Maldacena, D.~Stanford, and Z.~Yang, ``Diving into traversable wormholes,''
  \href{http://dx.doi.org/10.1002/prop.201700034}{{\em Fortsch. Phys.}
  {\bfseries 65} (2017) 1700034}.

\bibitem{Larkin68}
A.~Larkin and Y.~N. Ovchinnikov, ``Quasiclassical method in the theory of
  superconductivity,'' {\em JETP} {\bfseries 28} (1968) 1200.

\bibitem{shenker2014black}
S.~H. Shenker and D.~Stanford, ``Black holes and the butterfly effect,'' {\em
  JHEP} {\bfseries 3} (2014) 67.

\bibitem{Roberts:2015aa}
D.~A. Roberts, D.~Stanford, and L.~Susskind, ``Localized shocks,''
  \href{http://dx.doi.org/10.1007/JHEP03(2015)051}{{\em JHEP} {\bfseries 3}
  (2015) 51}.

\bibitem{Roberts:2017aa}
D.~A. Roberts and B.~Yoshida, ``Chaos and complexity by design,''
  \href{http://dx.doi.org/10.1007/JHEP04(2017)121}{{\em JHEP} {\bfseries 4}
  (2017) 121}.

\bibitem{Beigi:2013aa}
S.~Beigi, ``Sandwiched r{\'e}nyi divergence satisfies data processing
  inequality,'' \href{http://dx.doi.org/10.1063/1.4838855}{{\em J. Math. Phys.}
  {\bfseries 54} (2013) 122202}.

\bibitem{Yoshida:2019aa}
B.~Yoshida and N.~Y. Yao, ``Disentangling scrambling and decoherence via
  quantum teleportation,''
  \href{http://dx.doi.org/10.1103/PhysRevX.9.011006}{{\em Phys. Rev. X}
  {\bfseries 9} (2019) 011006--}.

\bibitem{Sekino08}
Y.~Sekino and L.~Susskind, ``Fast scramblers,'' {\em JHEP} {\bfseries 10}
  (2008) 065.

\bibitem{Shor18}
P.~W. Shor, ``Scrambling time and causal structure of the photon sphere of a
  schwarzschild black hole.'' Arxiv:1807.04363.

\bibitem{Dray:1985aa}
T.~Dray and G.~'t~Hooft, ``The effect of spherical shells of matter on the
  schwarzschild black hole,'' {\em Comm. Math. Phys.} {\bfseries 99} (1985)
  613--625.

\bibitem{Hooft:1987aa}
G.~'t~Hooft, ``Strings from gravity,'' {\em Physica Scripta} {\bfseries 15}
  (1987) 143.

\bibitem{Kiem:1995aa}
Y.~Kiem, H.~Verlinde, and E.~Verlinde, ``Black hole horizons and
  complementarity,'' \href{http://dx.doi.org/10.1103/PhysRevD.52.7053}{{\em
  Phys. Rev. D} {\bfseries 52} (1995) 7053--7065}.

\bibitem{Sfetsos:1995aa}
K.~Sfetsos, ``On gravitational shock waves in curved spacetimes,''
  \href{http://dx.doi.org/https://doi.org/10.1016/0550-3213(94)00573-W}{{\em
  Nucl. Phys. B} {\bfseries 436} (1995) 721--745}.

\bibitem{BenTov:2019aa}
Y.~BenTov and J.~Swearngin, ``Gravitational shockwaves on rotating black
  holes,'' \href{http://dx.doi.org/10.1007/s10714-019-2512-7}{{\em Gen. Rel.
  Grav.} {\bfseries 51} (2019) 25}.

\bibitem{Kitaev:2018aa}
A.~Kitaev and S.~J. Suh, ``The soft mode in the sachdev-ye-kitaev model and its
  gravity dual,'' \href{http://dx.doi.org/10.1007/JHEP05(2018)183}{{\em JHEP}
  {\bfseries 5} (2018) 183}.

\bibitem{Gu18}
Y.~Gu and A.~Kitaev, ``On the relation between the magnitude and exponent of
  otocs.'' Arxiv:1812.00120.

\bibitem{Srednicki:1994aa}
M.~Srednicki, ``Chaos and quantum thermalization,''
  \href{http://dx.doi.org/10.1103/PhysRevE.50.888}{{\em Phys. Rev. E}
  {\bfseries 50} (1994) 888--901}.

\bibitem{Ryu06}
S.~Ryu and T.~Takayanagi, ``Holographic derivation of entanglement entropy from
  the anti--de sitter space/conformal field theory correspondence,'' {\em Phys.
  Rev. Lett.} {\bfseries 96} (2006) 181602.

\bibitem{Harlow:2016aa}
D.~Harlow, ``Wormholes, emergent gauge fields, and the weak gravity
  conjecture,'' \href{http://dx.doi.org/10.1007/JHEP01(2016)122}{{\em JHEP}
  {\bfseries 1} (2016) 122}.

\bibitem{Bao:2018aa}
N.~Bao, S.~M. Carroll, A.~Chatwin-Davies, J.~Pollack, and G.~N. Remmen,
  ``Branches of the black hole wave function need not contain firewalls,''
  \href{http://dx.doi.org/10.1103/PhysRevD.97.126014}{{\em Phys. Rev. D}
  {\bfseries 97} (2018) 126014--}.

\bibitem{Papadodimas:2015aa}
K.~Papadodimas and S.~Raju, ``Local operators in the eternal black hole,''
  \href{http://dx.doi.org/10.1103/PhysRevLett.115.211601}{{\em Phys. Rev.
  Lett.} {\bfseries 115} (2015) 211601--}.

\end{thebibliography}
%
\end{document}